\documentclass[preprint,onecolumn,nofootinbib]{revtex4}
\pdfoutput=1
\usepackage[colorlinks=true,linkcolor=blue,urlcolor=blue,filecolor=black,citecolor=red,pdfstartview=FitV,pdftitle={},pdfsubject={},pdfkeywords={},pdfpagemode=None,bookmarksopen=true]{hyperref}
\usepackage{graphicx}
\usepackage{amsmath}
\usepackage{amsfonts}
\usepackage{amssymb}
\usepackage{color}%
\usepackage{dcolumn}
\newcommand{\f}{\begin{equation}}
\newcommand{\ff}{\end{equation}}
\newcommand{\fa}{\begin{eqnarray}}
\newcommand{\ffa}{\end{eqnarray}}

\usepackage{ulem}

\begin{document}
\title{Quasi-normal modes of holographic system with Weyl correction and momentum dissipation}
\author{Jian-Pin Wu $^{1,2}$}
\thanks{jianpinwu@mail.bnu.edu.cn}
\author{Peng Liu $^{3}$}
\thanks{phylp@jnu.edu.cn}
\affiliation{
$^1$~Center for Gravitation and Cosmology, College of Physical Science and Technology, Yangzhou University, Yangzhou 225009, China
\
\\
$^2$ Institute of Gravitation and Cosmology, Department of
Physics, School of Mathematics and Physics, Bohai University, Jinzhou 121013, China
\
\\
$^3$ Department of
Physics, Jinan University, Guangzhou 510632, China}

\begin{abstract}

We study the charge response in complex frequency plane and the quasi-normal modes (QNMs) of the boundary quantum field theory with momentum dissipation
dual to a probe generalized Maxwell system with Weyl correction.
When the strength of the momentum dissipation $\hat{\alpha}$ is small,
the pole structure of the conductivity is similar to the case without the momentum dissipation.
The qualitative correspondence between the poles of the real part of the conductivity of the original theory
and the ones of its electromagnetic (EM) dual theory approximately holds
when $\gamma\rightarrow -\gamma$ with $\gamma$ being the Weyl coupling parameter.
While the strong momentum dissipation alters the pole structure such that
most of the poles locate at the purely imaginary axis.
At this moment, the correspondence between the poles of the original theory and its EM dual one is violated when $\gamma\rightarrow -\gamma$.
In addition, for the dominant pole, the EM duality almost holds when $\gamma\rightarrow -\gamma$ for all $\hat{\alpha}$ except for
a small region of $\hat{\alpha}$.

\end{abstract}
\maketitle
\section{Introduction}

The quantum critical (QC) dynamics is a long-standing important issues in strongly coupling condensed matter systems,
in which the perturbative techniques in traditional field theory unfortunately lose its power.
An alternative method is the AdS/CFT correspondence \cite{Maldacena:1997re,Gubser:1998bc,Witten:1998qj,Aharony:1999ti},
which maps a strongly coupled quantum field
theory to a weakly coupled gravitational theory in the large N limit.
By this way, the holographic QC dynamics at zero density,
dual to a probe Maxwell field coupled to the Weyl tensor $C_{\mu\nu\rho\sigma}$ in the Schwarzschild-AdS (SS-AdS) in bulk,
has been widely studied in \cite{Myers:2010pk,Ritz:2008kh,WitczakKrempa:2012gn,WitczakKrempa:2013ht,Witczak-Krempa:2013nua,Katz:2014rla,Sachdev:2011wg,Hartnoll:2016apf}.
Since the Weyl tensor is taken into account, it exhibits non-trivial frequency dependent conductivity.
In particular, depending on the sign of the coupling parameter $\gamma$,
a Damle-Sachdev (DS) peak \cite{Damle:1997rxu} resembling the particle response
or a dip resembling the vortex response is observed in optical conductivity\footnote{Those kind of peak and dip features have also been observed in probe branes setups \cite{Chen:2017dsy}
and just higher terms in $F^2$ with $F$ being the Maxwell field strength \cite{Baggioli:2016oju}. Although the role and the presence of a Drude-like peak in the
optical conductivity have been partially explained for the probe brane case \cite{Chen:2017dsy}, it is still not so clear why there is a Drude-like peak and to what it is connected.
More efforts and attempts are deserved to further pursuit in future.} \cite{Myers:2010pk}.
It is analogous to that of the superfluid-insulator quantum critical point (QCP) \cite{Myers:2010pk,Sachdev:2011wg,Hartnoll:2016apf}.
Further, the charge response from higher derivative (HD) theory is studied,
in which we have an arbitrarily sharp Drude-like peak
or vanishing DC conductivity depending on the coupling parameter \cite{Witczak-Krempa:2013aea}.
Of particular interest is that the behavior of the holographic HD system is
very similar to the $O(N)$ $NL\sigma M$ model for large-$N$ \cite{Damle:1997rxu}.
Another important progress is the construction of neutral scalar hair black brane in bulk by introducing the coupling between Weyl tensor and neutral scalar field,
which is dual to QC dynamics and the one away from QCP in the boundary field theory \cite{Myers:2016wsu,Lucas:2017dqa}.

Also, we can introduce the momentum dissipation, implemented by a pair of axionic fields linearly dependent on spatial coordinates \cite{Andrade:2013gsa},
into the holographic QC systems studied in \cite{Myers:2010pk,Ritz:2008kh,WitczakKrempa:2012gn,WitczakKrempa:2013ht,Witczak-Krempa:2013nua,Katz:2014rla,Sachdev:2011wg},
which is away from QCP, to explore the corresponding effects.
We observe that for the $4$ derivative theory, the momentum dissipation drives the peak and dip into each other \cite{Wu:2016jjd},
while for the $6$ derivative theory, similar behaviors are not observed \cite{Fu:2017oqa}.
Another appealing phenomena for $4$ derivative theory is that there is a specific value of the momentum dissipation strength, i.e., $\hat{\alpha}=2/\sqrt{3}$,
for which the particle-vortex duality exactly remains \cite{Wu:2016jjd}.
In this paper, we want to extend the previous works \cite{Wu:2016jjd}
to the charge response in complex frequency plane.
We shall particularly focus on the properties of the quasi-normal modes (QNMs),
which corresponds to the poles of the retarded Green's function for the dual boundary CFT
(see \cite{Kovtun:2005ev,Berti:2009kk,Konoplya:2011qq,Horowitz:1999jd,Wang:2004bv,Wang:2000gsa,Giammatteo:2004wp,Yao:2011kf,Lin:2016sch,Gursoy:2016ggq,Gursoy:2016tgf,Jansen:2017oag,Cardoso:2017soq} and references therein).
Our paper is organized as follows.
In Section \ref{sec-HF}, we introduce the holographic framework, including
the Einstein-axions (EA) theory, which is the background geometry dual to a specific thermal excited state with momentum dissipation,
and the $4$ derivative theory without electromagnetic (EM) duality, which is regarded as the probe on top of the EA background geometry.
And then, we calculate the conductivity in the complex frequency plane in Section \ref{sec-cond}.
We are in particular interested in the QNMs of our present models, which are presented in Section \ref{sec-QNM}.
The conclusions and discussions are summarized in Section \ref{sec-conclusion}.

\section{Holographic framework}\label{sec-HF}

We consider a specific thermal state with homogeneous disorder,
which is holographically described by the EA theory \cite{Andrade:2013gsa} (also refer to \cite{Davison:2014lua,Grozdanov:2015qia,Baggioli:2014roa,Kuang:2017cgt,Ge:2014aza,Ling:2016dck,Kim:2014bza} and references therein)
\fa
\label{ac-ax}
S_0=\int d^4x\sqrt{-g}\Big(R+6-\frac{1}{2}\sum_{I=x,y}(\partial \phi_I)^2\Big)
\,,
\ffa
where $\phi_I=\alpha x_I$ with $I=x,y$ and $\alpha$ being a constant.
In this action, there is a negative cosmological constant $\Lambda=-6$, which supports an asymptotically AdS spacetimes \footnote{Here,
without loss of generality we have set the AdS radius $L=1$ for simplicity.}.
The EA action (\ref{ac-ax}) gives
a neutral black brane solution \cite{Andrade:2013gsa}
\fa
\label{bl-br}
&&
ds^2=\frac{1}{u^2}\Big(-f(u)dt^2+\frac{1}{f(u)}du^2+dx^2+dy^2\Big)\,,
\nonumber
\\
&&
f(u)=(1-u)p(u)\,,~~~~~~~
p(u)=\frac{\sqrt{1+6\hat{\alpha}^2}-2\hat{\alpha}^2-1}{\hat{\alpha}^2}u^2+u+1\,.
\ffa
$u=0$ is the asymptotically AdS boundary while the horizon locates at $u=1$.
Here we have parameterized the black brane solution by $\hat{\alpha}=\alpha/4\pi T$
with the Hawking temperature $T=p(1)/4\pi$.
Note that for the particular way we adopt to break translations, i.e., the axionic fields,
the original (spacetime translations) $\times$ (internal translations) group is broken to the diagonal subgroup such that the geometry is homogeneous \cite{Nicolis:2015sra,Alberte:2015isw}.

On top of the geometry background (\ref{bl-br}), we consider the following $4$ derivative theory \cite{Myers:2010pk}
(also see \cite{WitczakKrempa:2012gn,WitczakKrempa:2013ht,Witczak-Krempa:2013nua,Witczak-Krempa:2013aea,Katz:2014rla,Wu:2016jjd,Fu:2017oqa})
\fa
\label{ac-ma}
S_A=\int d^4x\sqrt{-g}\Big(-\frac{1}{8g_F^2}F_{\mu\nu}X^{\mu\nu\rho\sigma}F_{\rho\sigma}\Big)\,,
\ffa
where
\fa
X_{\mu\nu}^{\ \ \rho\sigma}=I_{\mu\nu}^{\ \ \rho\sigma}-8\gamma C_{\mu\nu}^{\ \ \rho\sigma}\,,
~~~~~~~
I_{\mu\nu}^{\ \ \rho\sigma}=\delta_{\mu}^{\ \rho}\delta_{\nu}^{\ \sigma}-\delta_{\mu}^{\ \sigma}\delta_{\nu}^{\ \rho}\,.
\label{X-tensor}
\ffa
The new tensor $X$ in the above equation possess the following symmetries
\fa
X_{\mu\nu\rho\sigma}=X_{[\mu\nu][\rho\sigma]}=X_{\rho\sigma\mu\nu}\,.
\label{X-sym}
\ffa
When we set $X_{\mu\nu}^{\ \ \rho\sigma}=I_{\mu\nu}^{\ \ \rho\sigma}$, the generalized Maxwell theory (\ref{ac-ma}) reduces to the standard version.
And then, we can write down the equation of motion from the action (\ref{ac-ma}),
\fa
\nabla_{\nu}(X^{\mu\nu\rho\sigma}F_{\rho\sigma})=0\,.
\label{eom-Max}
\ffa

Next, we can construct the corresponding dual EM theory,
which is \cite{Myers:2010pk}
\fa
\label{ac-SB}
S_B=\int d^4x\sqrt{-g}\Big(-\frac{1}{8\hat{g}_F}G_{\mu\nu}\widehat{X}^{\mu\nu\rho\sigma}G_{\rho\sigma}\Big)\,,
\ffa
where $\hat{g}_F^2\equiv 1/g_F^2$ and $G_{\mu\nu}\equiv\partial_{\mu}B_{\nu}-\partial_{\nu}B_{\mu}$ is the dual field strength.
The tensor $\widehat{X}$ satisfies
\fa
\widehat{X}_{\mu\nu}^{\ \ \rho\sigma}=-\frac{1}{4}\varepsilon_{\mu\nu}^{\ \ \alpha\beta}(X^{-1})_{\alpha\beta}^{\ \ \gamma\lambda}\varepsilon_{\gamma\lambda}^{\ \ \rho\sigma}\,,
~~~~~~
\frac{1}{2}(X^{-1})_{\mu\nu}^{\ \ \rho\sigma}X_{\rho\sigma}^{\ \ \alpha\beta}\equiv I_{\mu\nu}^{\ \ \alpha\beta}\,.
\label{X-hat}
\ffa
And then the equation of motion of the dual theory (\ref{ac-SB}) can be written as
\fa
\nabla_{\nu}(\widehat{X}^{\mu\nu\rho\sigma}G_{\rho\sigma})=0\,.
\label{eom-Max-B}
\ffa

For the standard Maxwell theory in four dimensional bulk spacetimes, $\widehat{X}_{\mu\nu}^{\ \ \rho\sigma}=I_{\mu\nu}^{\ \ \rho\sigma}$,
which indicates that both the theories (\ref{ac-ma}) and (\ref{ac-SB})
are identical and so the Maxwell theory is self-dual.
The coupling term between the Weyl tensor and the Maxwell field strength breaks such self-duality.
But for small $\gamma$, since
\fa
&&
(X^{-1})_{\mu\nu}^{\ \ \rho\sigma}=I_{\mu\nu}^{\ \ \rho\sigma}+8\gamma C_{\mu\nu}^{\ \ \rho\sigma}+\mathcal{O}(\gamma^2)\,,
\label{Xin}
\\
&&
\widehat{X}_{\mu\nu}^{\ \ \rho\sigma}=(X^{-1})_{\mu\nu}^{\ \ \rho\sigma}+\mathcal{O}(\gamma^2)\,ㄛ
\ffa
we have an approximate duality between the theories (\ref{ac-ma}) and (\ref{ac-SB})
with the change of the sign of $\gamma$.

\section{Conductivity in the complex frequency plane}\label{sec-cond}
\begin{figure}
\center{
\includegraphics[scale=0.35]{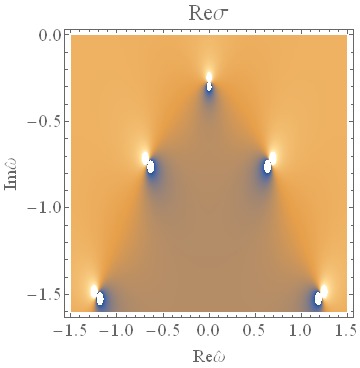}\ \hspace{0.8cm}
\includegraphics[scale=0.35]{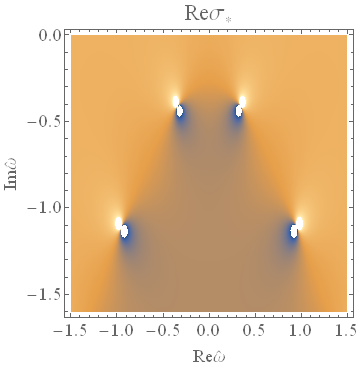}\ \\
\includegraphics[scale=0.35]{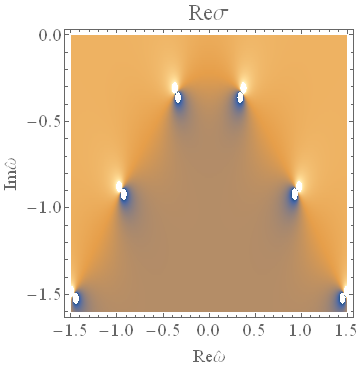}\ \hspace{0.8cm}
\includegraphics[scale=0.35]{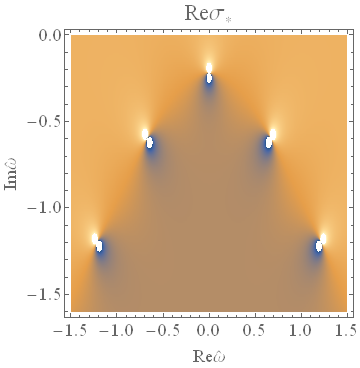}\ \\
\caption{\label{fig-cond-CF-alphah-0} Conductivity $\sigma$ and its dual $\sigma_{*}$ in the LHP for
$|\gamma|=1/12$ (the panels above are for $\gamma=1/12$ and the ones below for $\gamma=-1/12$)
and $\hat{\alpha}=0$ in the complex frequency plane. }}
\end{figure}
\begin{figure}
\center{
\includegraphics[scale=0.35]{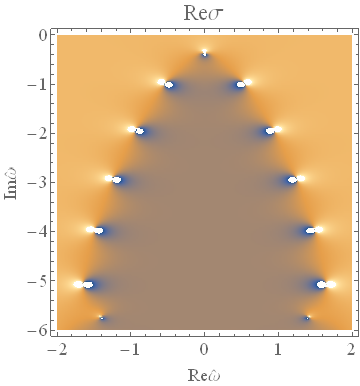}\ \hspace{0.8cm}
\includegraphics[scale=0.35]{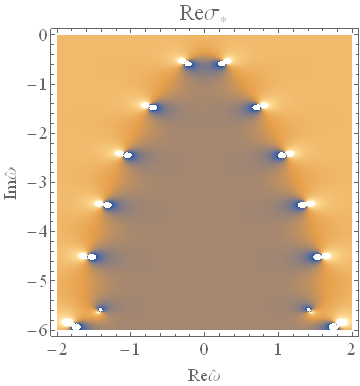}\ \\
\includegraphics[scale=0.35]{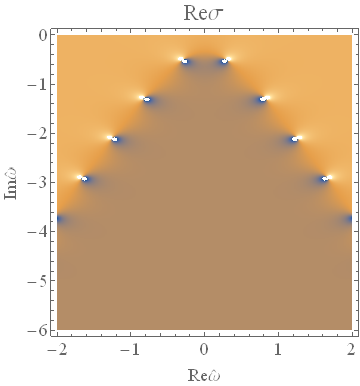}\ \hspace{0.8cm}
\includegraphics[scale=0.35]{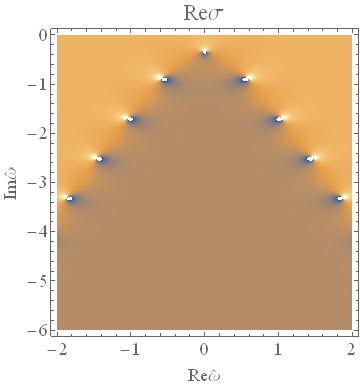}\ \\
\caption{\label{fig-cond-CF-alphah-0p5} Conductivity $\sigma$ and its dual $\sigma_{*}$ in the LHP for
$|\gamma|=1/12$ (the panels above are for $\gamma=1/12$ and the ones below for $\gamma=-1/12$)
and $\hat{\alpha}=1/2$ in the complex frequency plane.}}
\end{figure}
In our previous works \cite{Wu:2016jjd}, the conductivity on the real frequency axis was numerically studied.
Here, we extend the study from real frequency axis to complex frequency plane,
$\omega\equiv \texttt{Re}\omega+i\texttt{Im}\omega\in \mathbb{C}$.
To this end, we turn on the perturbation of the gauge field along $y$ direction like
$
A_{y}(t,u)\sim e^{-i\omega t}A_{y}(u)
$
.
And then, the perturbative equation can be written down \cite{Myers:2010pk,Wu:2016jjd}
\fa
A''_y
+\Big(\frac{f'}{f}+\frac{X'_6}{X_6}\Big)A'_y
+\frac{\mathfrak{p}^2\hat{\omega}^2}{f^2}\frac{X_2}{X_6}A_y
=0\,,
\label{Ma-Ay}
\ffa
where $X_i$, $i=1,\ldots,6$, are the components of $X_{A}^{\ B}$ defined as
$
X_{A}^{\ B}=\{X_1(u),X_2(u),X_3(u),X_4(u),X_5(u),X_6(u)\}\,,
$
with
$
A,B\in\{tx,ty,tu,xy,xu,yu\}\,.
$
Note that $X_1(u)=X_2(u)$ and $X_5(u)=X_6(u)$ due to the isotropy.
In addition, the dimensionless frequency
$
\hat{\omega}\equiv\frac{\omega}{4\pi T}=\frac{\omega}{\mathfrak{p}}\,,
$
with
$
\mathfrak{p}\equiv p(1)=4\pi T
$,
has been introduced in the perturbative equation (\ref{Ma-Ay}).
Next, we can numerically solve the Eq.(\ref{Ma-Ay}) with the ingoing condition at horizon on top of complex frequency plan
and read off the optical conductivity in terms of
\fa
\sigma(\omega)=\frac{\partial_uA_y(u,\hat{\omega})}{i\omega A_y(u,\hat{\omega})}\,.
\label{con-def}
\ffa
By letting $A_{\mu}\rightarrow B_{\mu}$ and $X_i\rightarrow \widehat{X}_i=1/X_i$, the corresponding equation of motion for the dual theory can be obtained.

\begin{figure}
\center{
\includegraphics[scale=0.35]{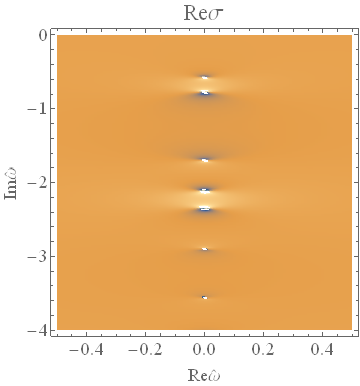}\ \hspace{0.8cm}
\includegraphics[scale=0.35]{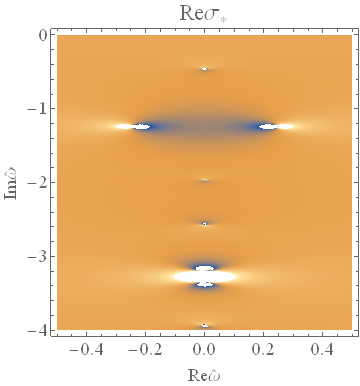}\ \\
\includegraphics[scale=0.35]{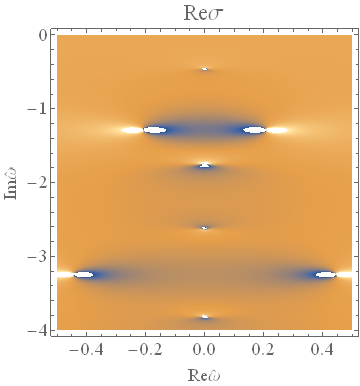}\ \hspace{0.8cm}
\includegraphics[scale=0.35]{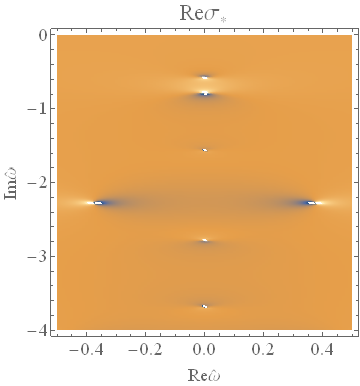}\ \\
\caption{\label{fig-cond-CF-alphah-2osqrt3} Conductivity $\sigma$ and its dual $\sigma_{*}$ in the LHP for
$|\gamma|=1/12$ (the panels above are for $\gamma=1/12$ and the ones below for $\gamma=-1/12$)
and $\hat{\alpha}=2/\sqrt{3}$ in the complex frequency plane.}}
\end{figure}
\begin{figure}
\center{
\includegraphics[scale=0.35]{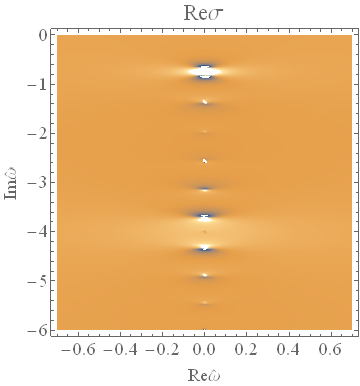}\ \hspace{0.8cm}
\includegraphics[scale=0.35]{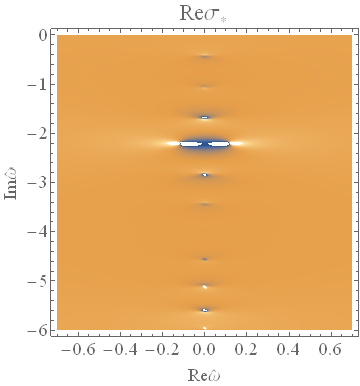}\ \\
\includegraphics[scale=0.35]{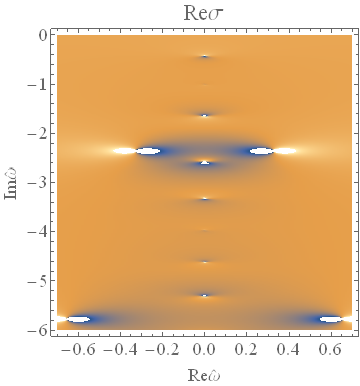}\ \hspace{0.8cm}
\includegraphics[scale=0.35]{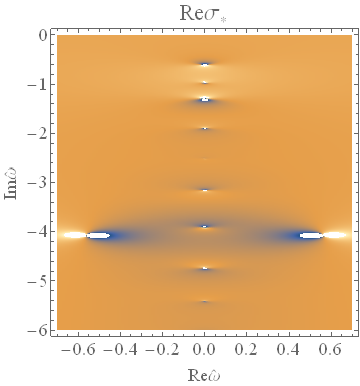}\ \\
\caption{\label{fig-cond-CF-alphah-2} Conductivity $\sigma$ and its dual $\sigma_{*}$ in the LHP for
$|\gamma|=1/12$ (the panels above are for $\gamma=1/12$ and the ones below for $\gamma=-1/12$)
and $\hat{\alpha}=2$ in the complex frequency plane.}}
\end{figure}

The numerical results are shown in FIG.\ref{fig-cond-CF-alphah-0}, \ref{fig-cond-CF-alphah-0p5}, \ref{fig-cond-CF-alphah-2osqrt3} and \ref{fig-cond-CF-alphah-2}
for the representative $\hat{\alpha}$ and the two values of $\gamma$ satisfying the bound, $\gamma=\pm 1/12$. We find that all the poles of the conductivity are in the lower-half plane (LHP),
which indicates that the perturbative modes are stable ones.
As a check on our numerics, we show the conductivity $\sigma$ (left panels)
and its dual $\sigma_{*}$ (right panels) in the LHP for
$|\gamma|=1/12$ (the panels above are for $\gamma=1/12$ and the ones below for $\gamma=-1/12$)
and $\hat{\alpha}=0$ in the complex frequency plane,
which have been obtained in \cite{WitczakKrempa:2012gn}.
Our results are in excellent agreement with that in \cite{WitczakKrempa:2012gn}.
For comparison, we firstly summary the main properties of the pole structure as what follows.
\begin{itemize}
  \item For $\gamma=1/12$, the dominant pole in $\texttt{Re}\sigma$, i.e., the pole closest to the real axis,
  is purely imaginary Drude-like pole which can be quantified as $\hat{\omega}=-i\Gamma$
with $\Gamma$ being the dissipation rate.
It corresponds to a particle-like transport
with a Drude-like peak at low frequency on the real frequency axis which has been shown in \cite{Myers:2010pk}.
\item While for $\gamma=-1/12$, the dominant poles in $\texttt{Re}\sigma$ are off-axis,
which corresponds to non-Drude-like transport with a dip at low frequency on the real frequency axis \cite{Myers:2010pk}.
\item As the increase of the frequency, some satellite poles in $\texttt{Re}\sigma$ emerge in pairs both for $\gamma=1/12$ and $\gamma=-1/12$,
which are off-axis and are obviously non-Drude-like.
\item There is a qualitative correspondence between the poles of $\texttt{Re}\sigma(\hat{\omega};\gamma)$
and the ones of $\texttt{Re}\sigma_{*}(\hat{\omega};-\gamma)$.
\end{itemize}

And then we explore the properties of the conductivity in the complex frequency plane
when the effect of the momentum dissipation is introduced.
The main properties are summarized as follows.
\begin{itemize}
  \item When the strength of the momentum dissipation is small, the pole structure of $\texttt{Re}\sigma$ is similar with that without the momentum dissipation (see FIG.\ref{fig-cond-CF-alphah-0p5} for $\hat{\alpha}=1/2$).
  And so qualitative correspondence between the poles of $\texttt{Re}\sigma(\hat{\omega};\gamma)$
and the ones of $\texttt{Re}\sigma_{*}(\hat{\omega};-\gamma)$ still holds for small $\hat{\alpha}$ (FIG.\ref{fig-cond-CF-alphah-0p5}),
which indicates we have a particle-vortex duality in the dual boundary field theory for small $\hat{\alpha}$.
\item As the strength of the momentum dissipation is increased, for example, $\hat{\alpha}=2/\sqrt{3}$ (FIG.\ref{fig-cond-CF-alphah-2osqrt3}) and $\hat{\alpha}=2$ (FIG.\ref{fig-cond-CF-alphah-2}),
  most of the poles in $\texttt{Re}\sigma$ and $\texttt{Re}\sigma_{*}$ locate at the purely imaginary axis both for $\gamma=1/12$ and $\gamma=-1/12$.
  But there are some exceptions in high frequency, which are off axis (FIG.\ref{fig-cond-CF-alphah-2osqrt3} and FIG.\ref{fig-cond-CF-alphah-2}).
  The particle-vortex duality is violated when the off axis modes in $\texttt{Re}\sigma$ or $\texttt{Re}\sigma_{*}$ appear.
\end{itemize}

\section{Quasi-normal modes}\label{sec-QNM}

QNMs are by definition the poles of the Green's function and are directly related to the conductivity.
Comparing with that of direct solution of conductivity on complex frequency,
the method of QNMs has a wider range of applicability and is numerically more stable and high efficient
such that we can carry out the quantitative study on the pole structure and obtain more insight.

To calculate the QNMs, we shall follow the method outlined in \cite{Jansen:2017oag}
and work in the advanced Eddington-Finkelstein coordinate, which is
\fa
ds^2=\frac{1}{u^2}\Big(-f(u)dt^2-2dtdu+dx^2+dy^2\Big)\,.
\label{dsAEF}
\ffa
And then, the perturbative equation (\ref{Ma-Ay}) can be changed correspondingly as
\fa
A_y' \left(\frac{f'}{f}+\frac{2 i \mathfrak{p} \hat{\omega} }{f}+\frac{X_6'}{X_6}\right)+\frac{i \mathfrak{p} \hat{\omega}
   A_y X_2'}{f X_6}+A_y''
   =0\,.
   \label{Ma-Ay-r}
\ffa
Again, the equation for the dual theory can be obtained by letting $A_{\mu}\rightarrow B_{\mu}$ and $X_i\rightarrow \widehat{X}_i=1/X_i$.
Imposing the ingoing boundary at the horizon, we solve the Eq. \eqref{Ma-Ay-r} to obtain QNMs.

\begin{figure}
\center{
\includegraphics[scale=0.28]{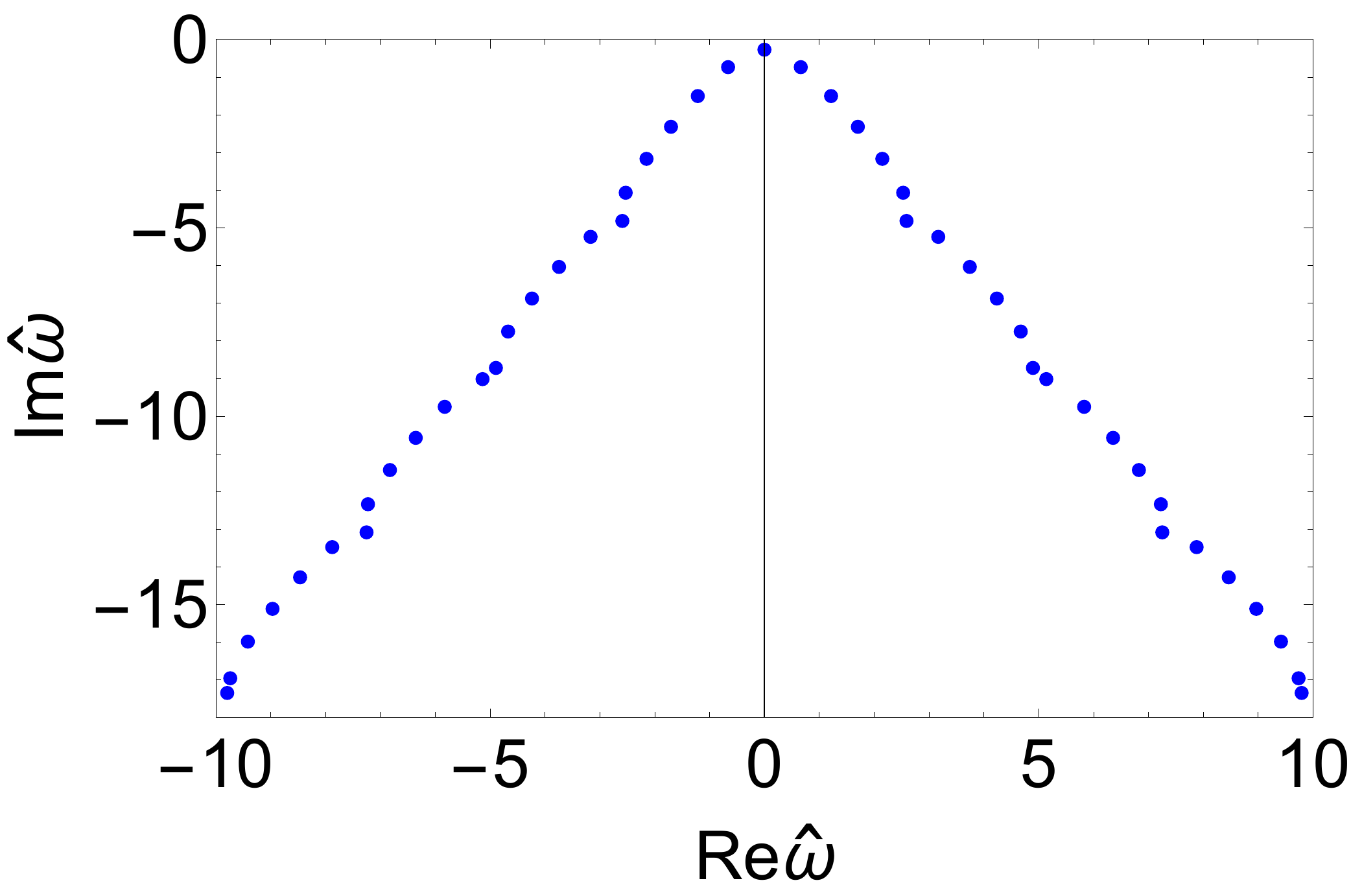}\ \hspace{0.5cm}
\includegraphics[scale=0.28]{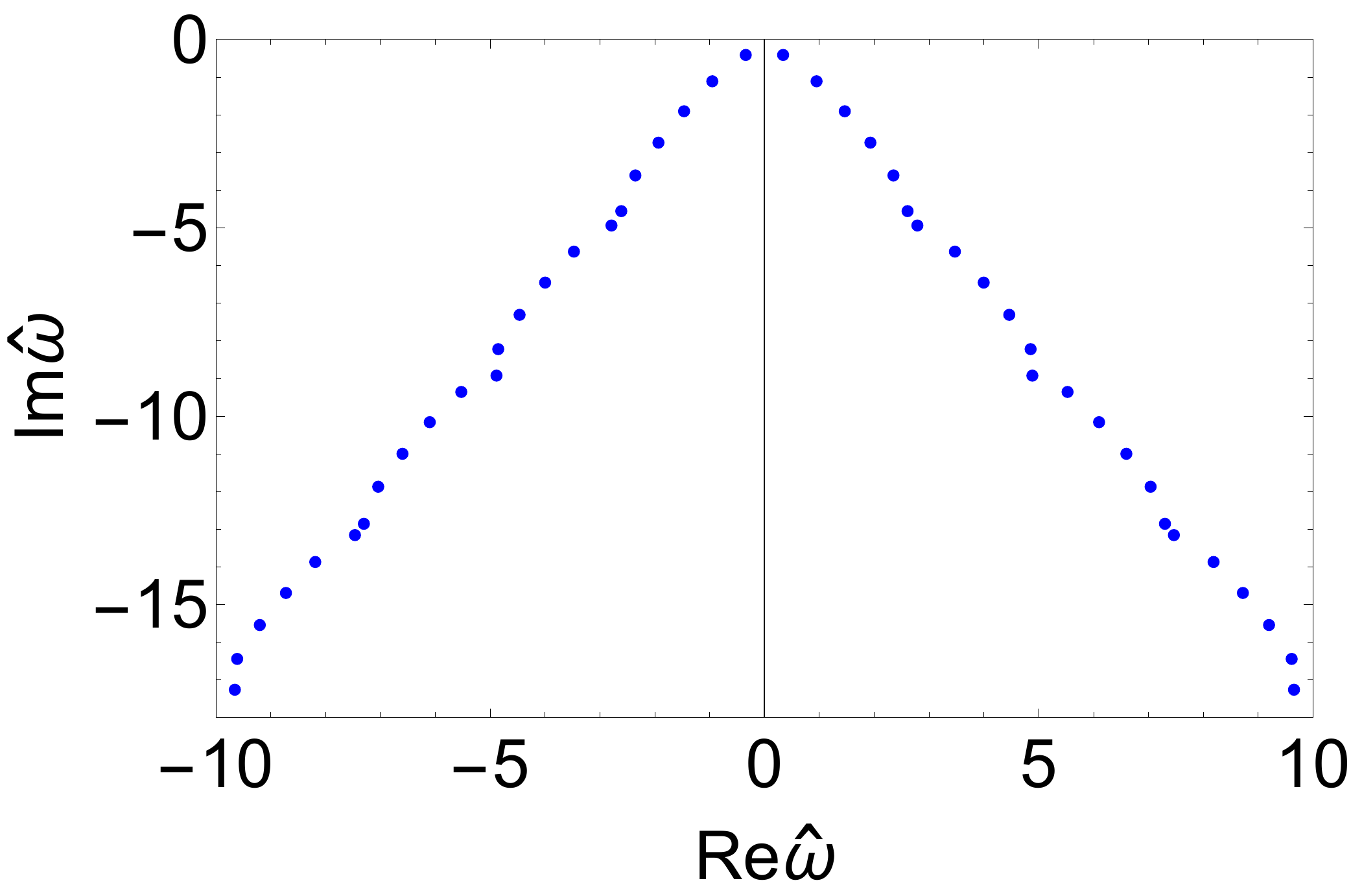}\ \\
\includegraphics[scale=0.28]{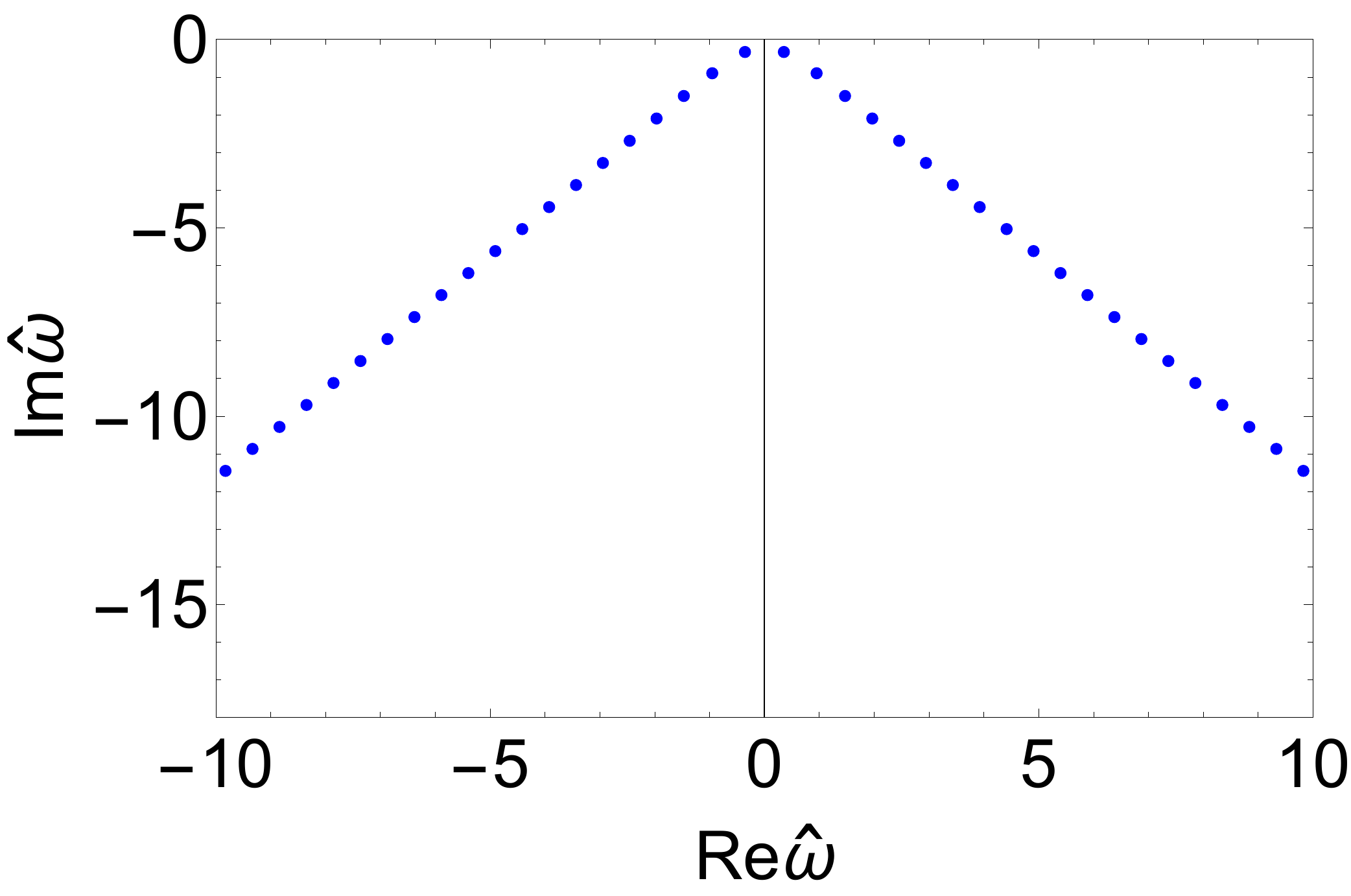}\ \hspace{0.5cm}
\includegraphics[scale=0.28]{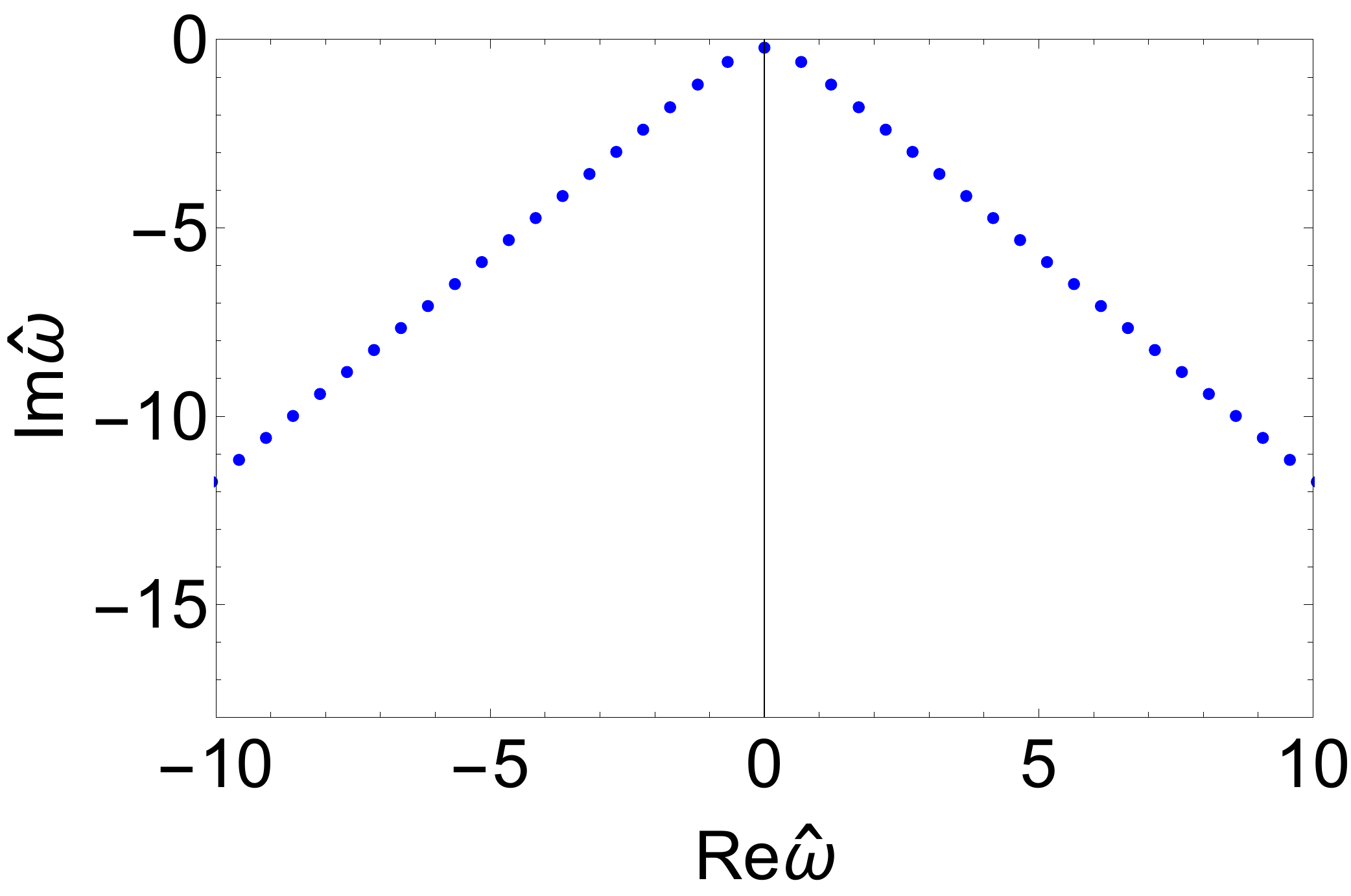}\ \\
\caption{\label{fig-QNM-alphah-0} QNMs (blue spots) of the transverse gauge mode for
$|\gamma|=1/12$ (the panels above are for $\gamma=1/12$ and the ones below for $\gamma=-1/12$)
and $\hat{\alpha}=0$ in the complex frequency plane.
The left panels are the QNMs of the gauge mode, and the right ones are that of the dual gauge mode.}}
\end{figure}
\begin{figure}
\center{
\includegraphics[scale=0.28]{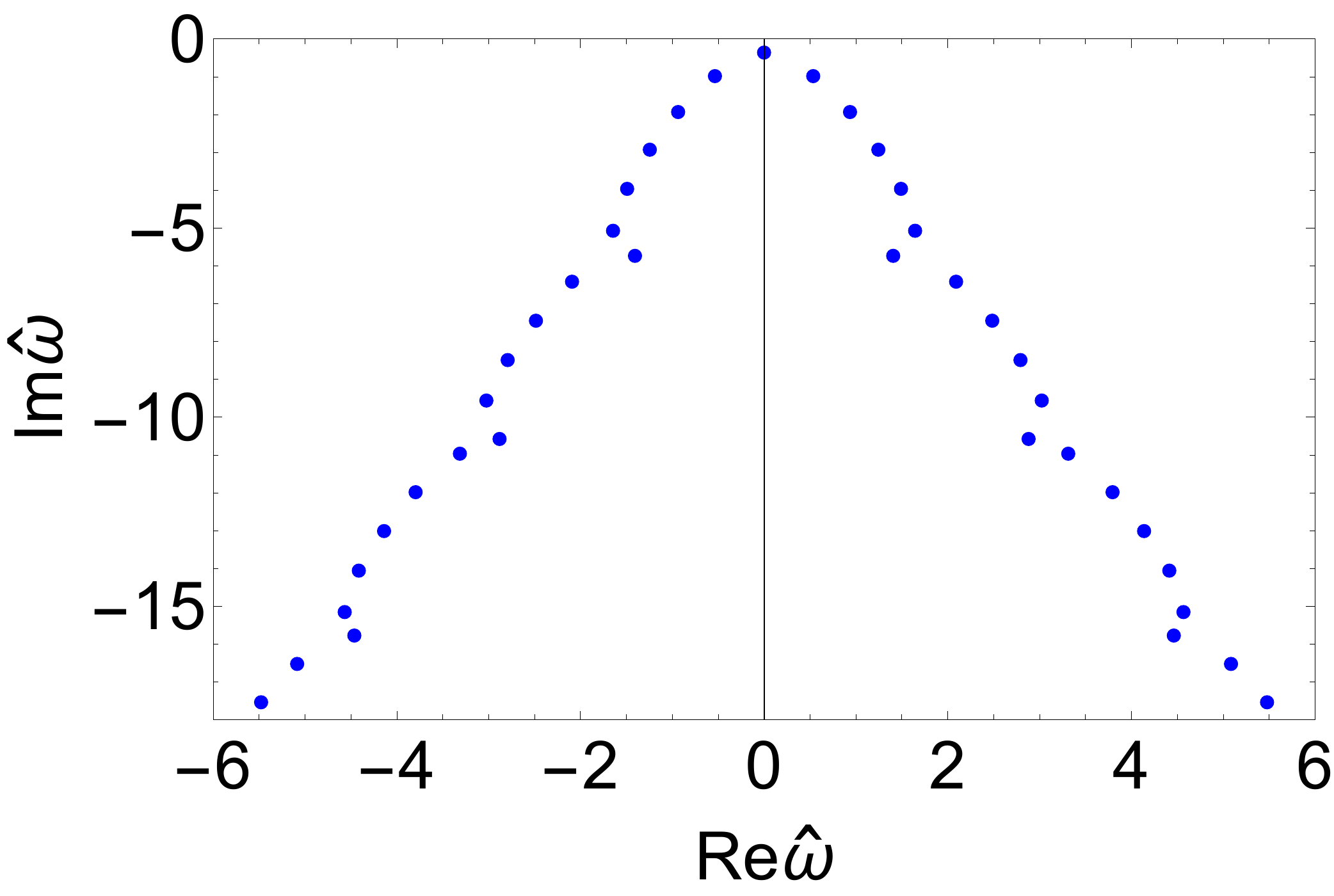}\ \hspace{0.5cm}
\includegraphics[scale=0.28]{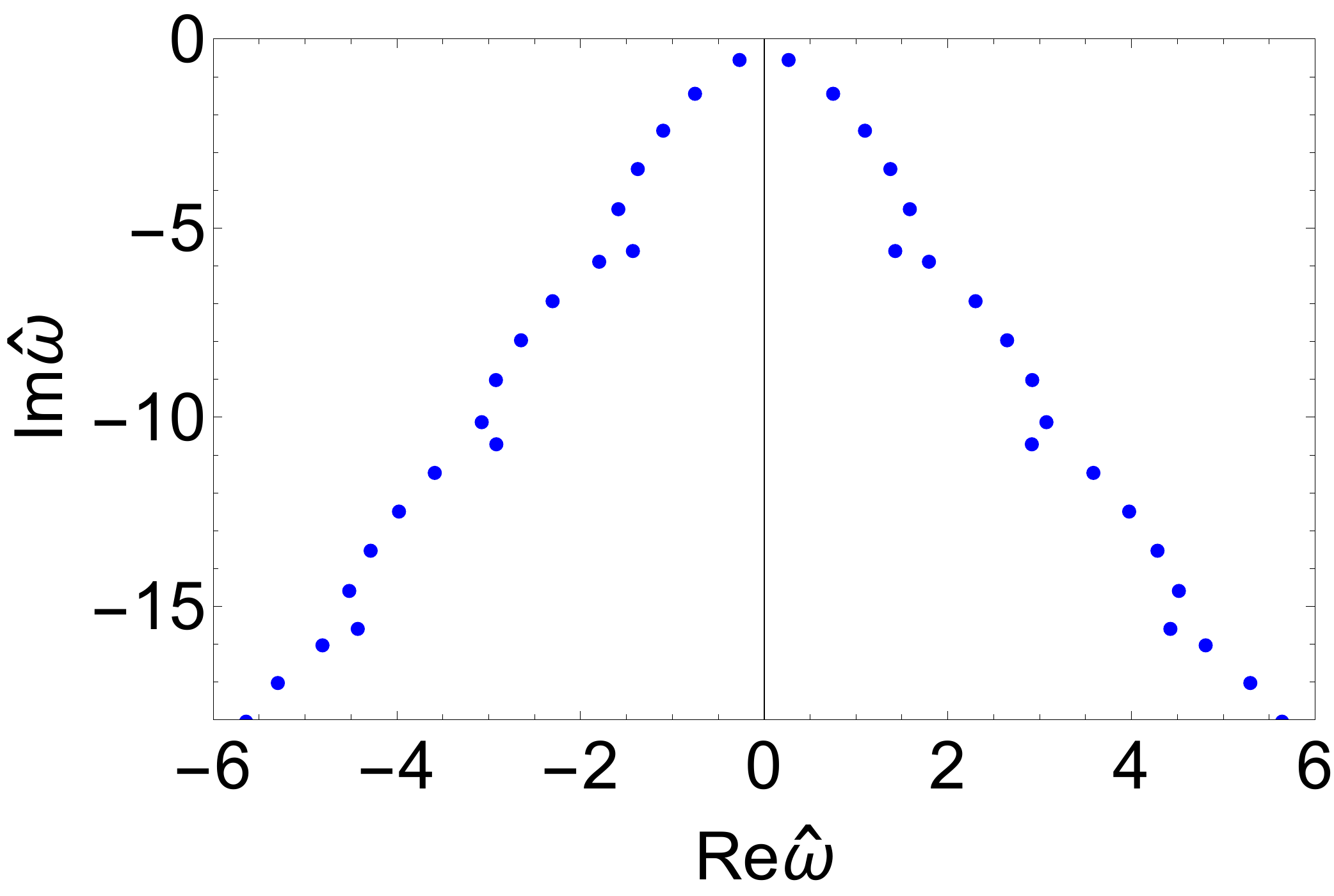}\ \\
\includegraphics[scale=0.28]{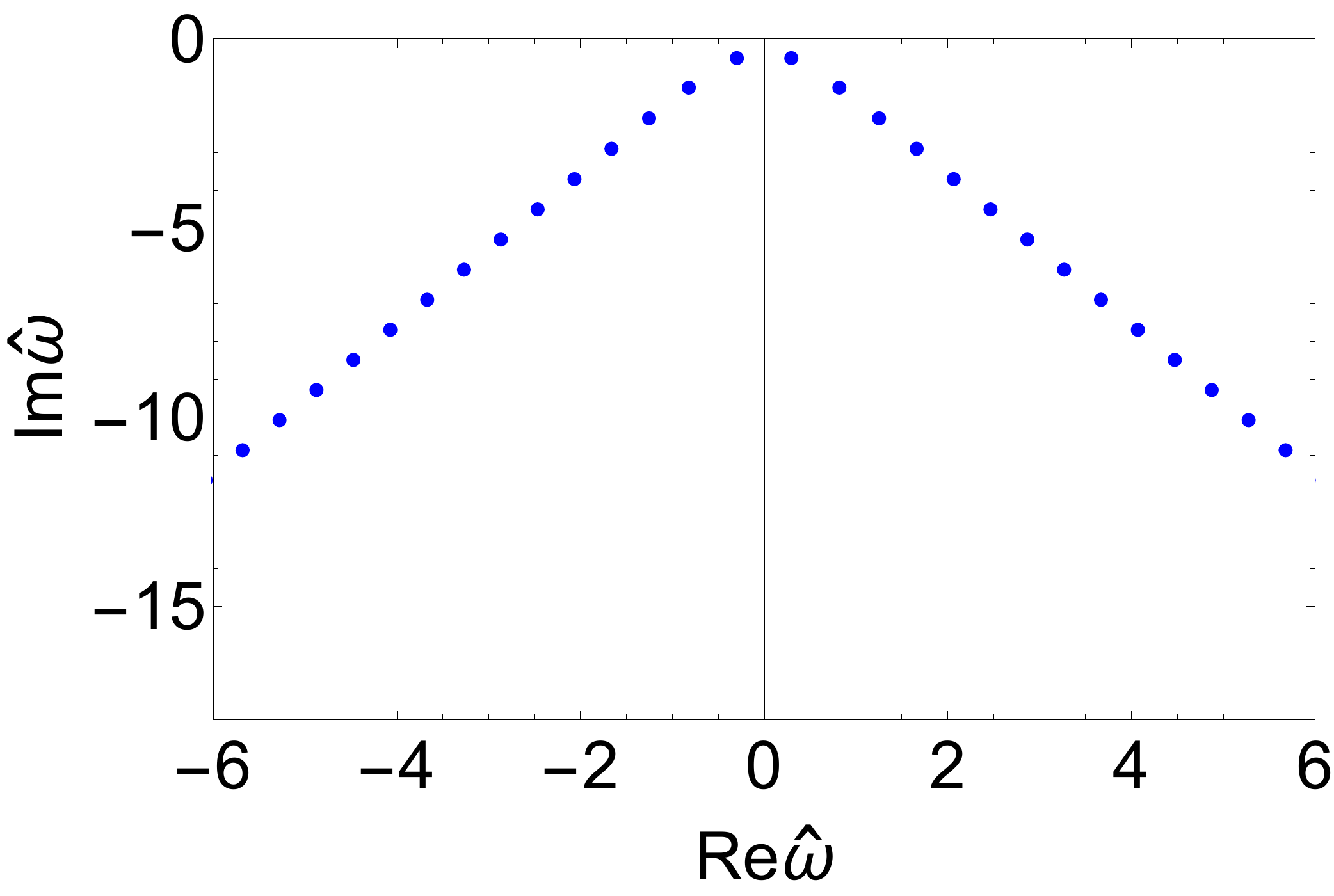}\ \hspace{0.5cm}
\includegraphics[scale=0.28]{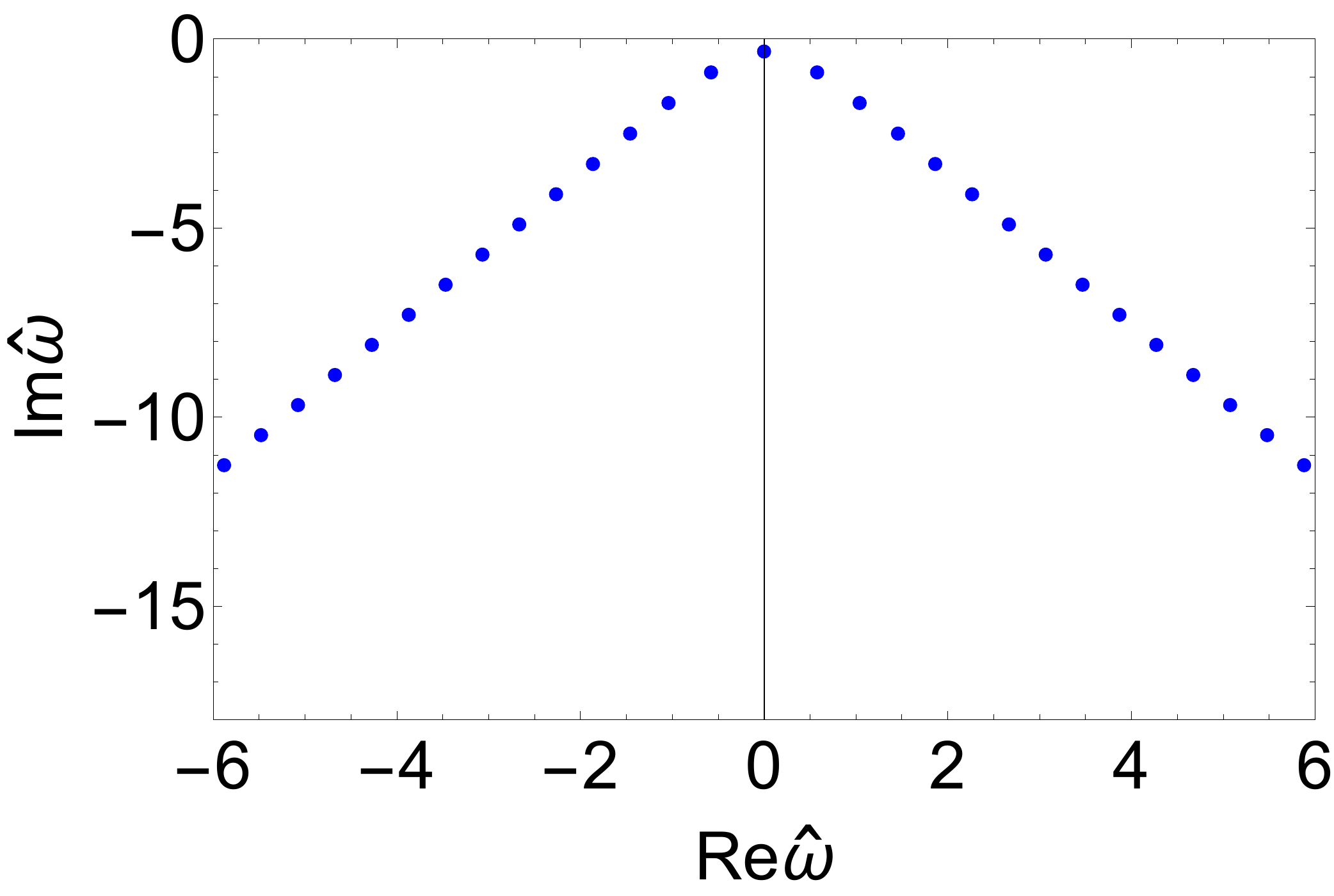}\ \\
\caption{\label{fig-QNM-alphah-1o2} QNMs (blue spots) of the transverse gauge mode for $|\gamma|=1/12$ (the panels above are for $\gamma=1/12$ and the ones below for $\gamma=-1/12$)
and $\hat{\alpha}=1/2$ in the complex frequency plane.
The left panels are the QNMs of the gauge mode, and the right ones are that of the dual gauge mode.}}
\end{figure}
\begin{figure}
\center{
\includegraphics[scale=0.28]{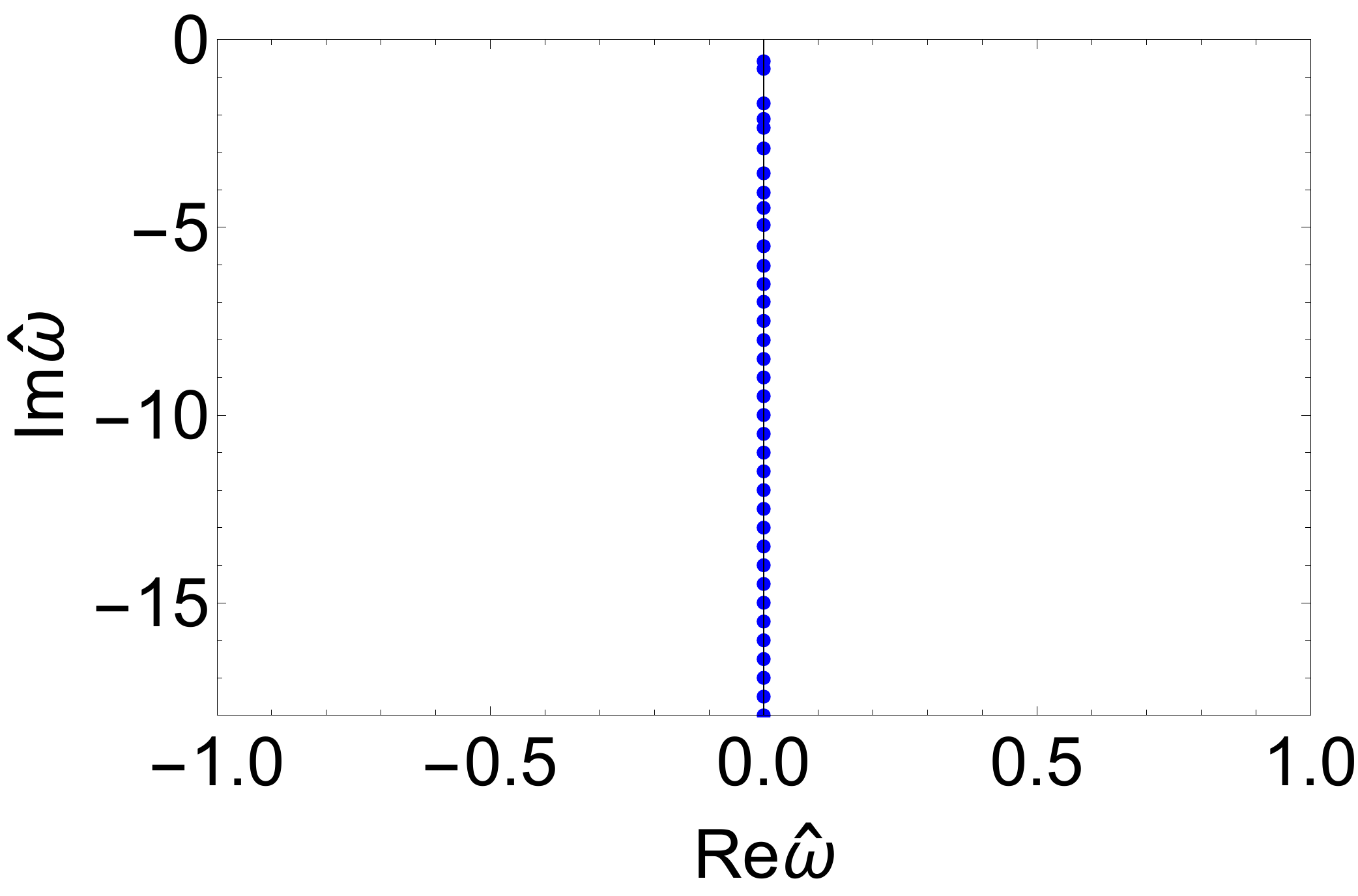}\ \hspace{0.5cm}
\includegraphics[scale=0.28]{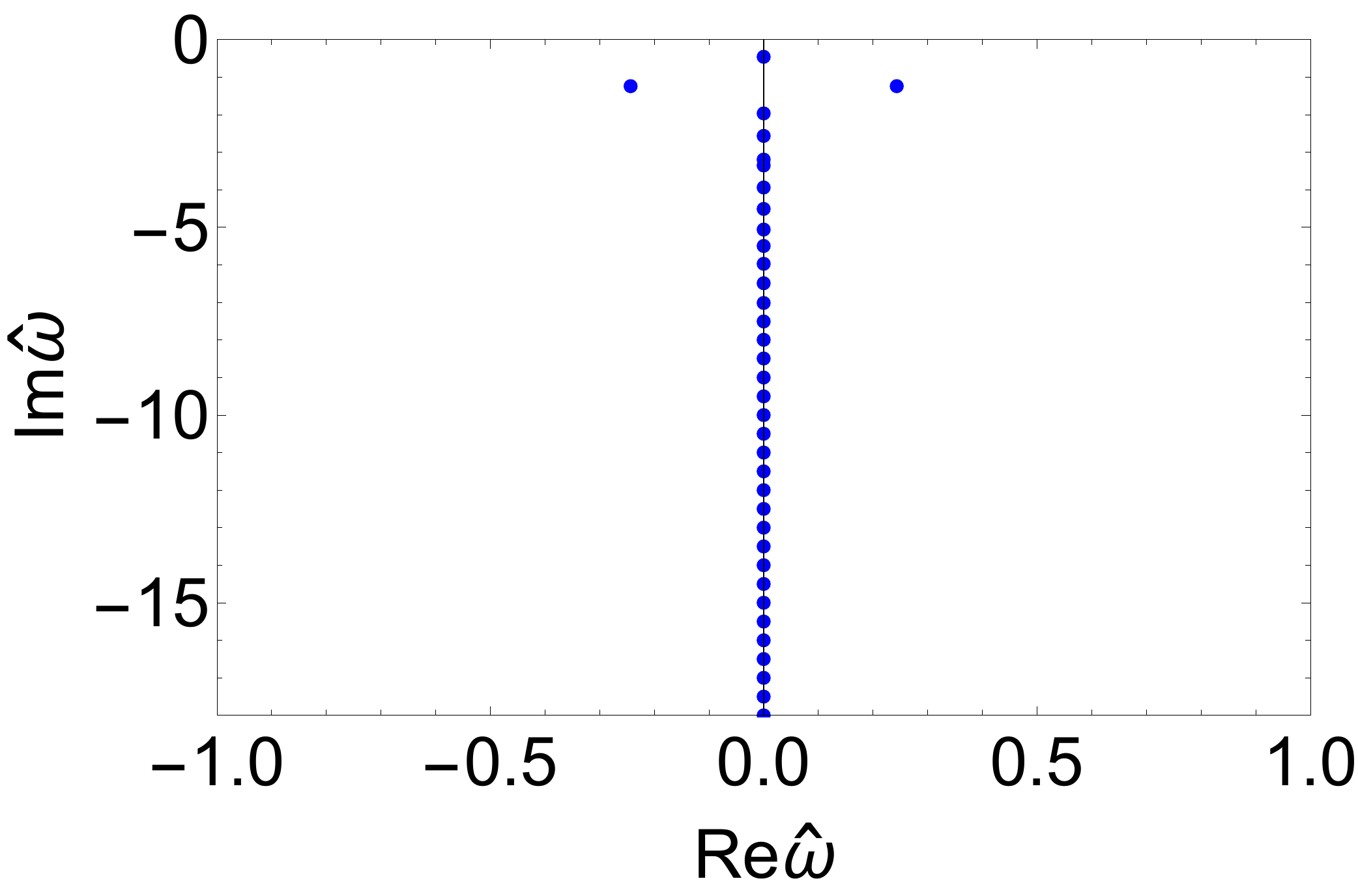}\ \\
\includegraphics[scale=0.28]{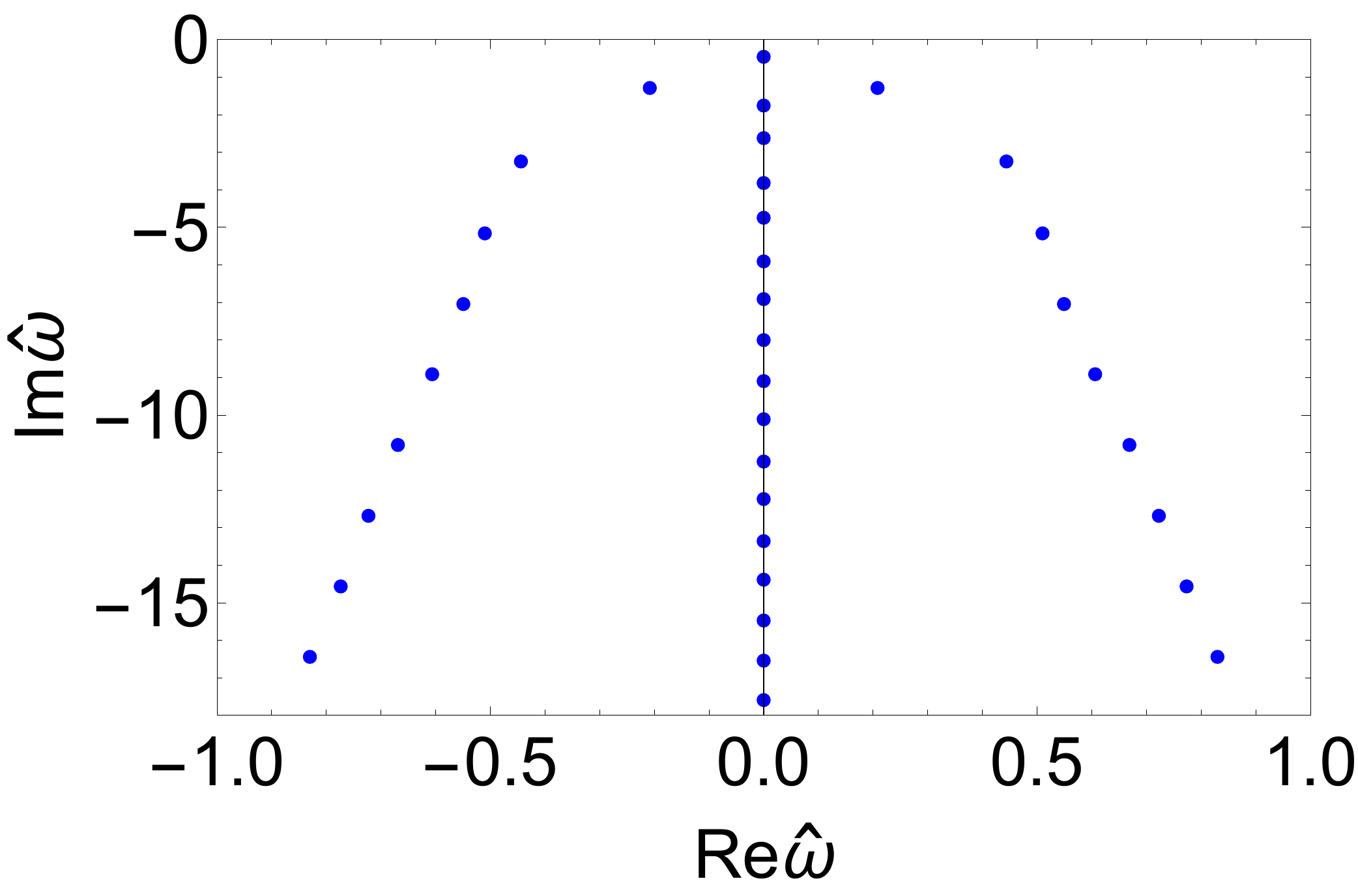}\ \hspace{0.5cm}
\includegraphics[scale=0.28]{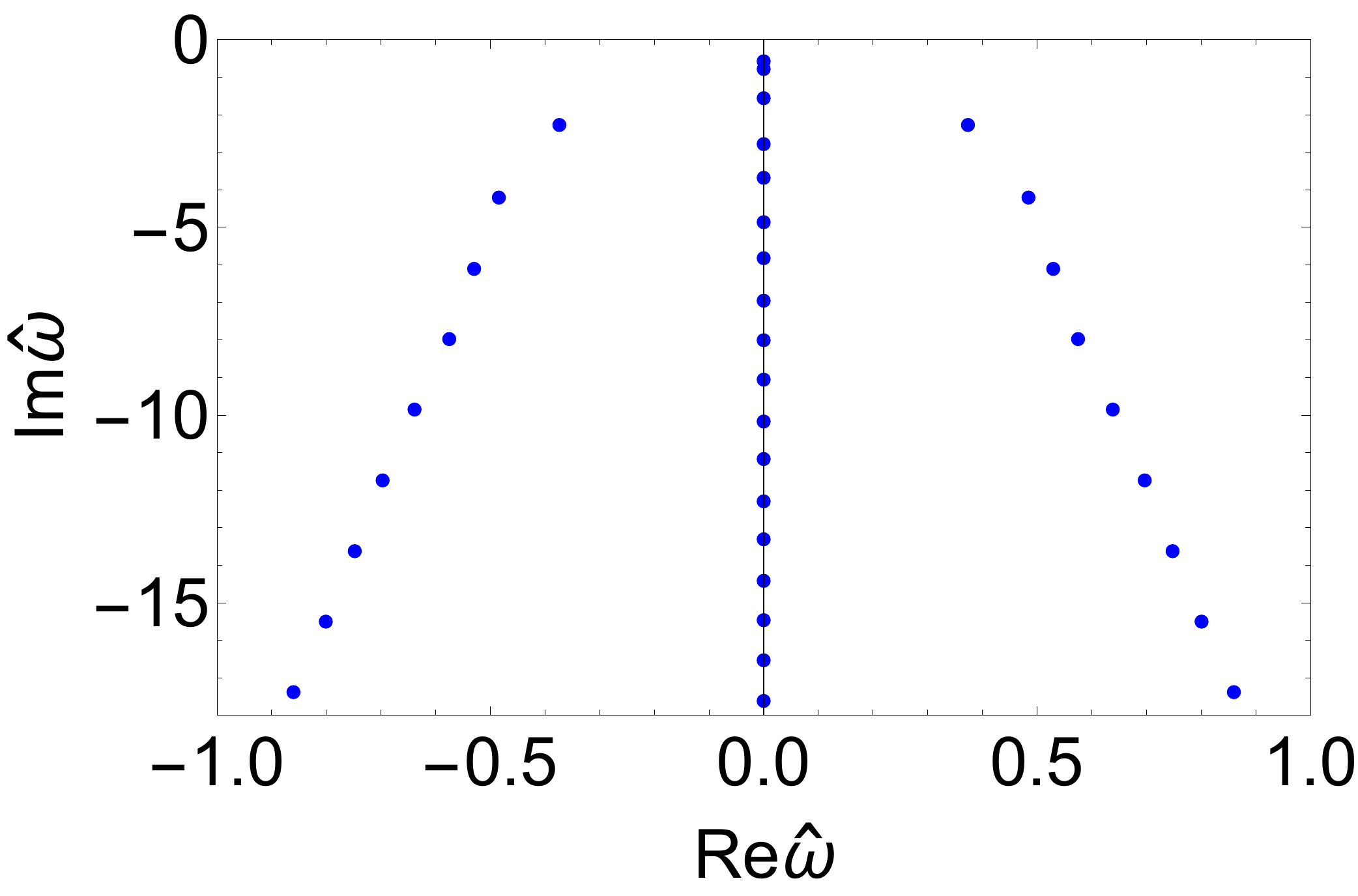}\ \\
\caption{\label{fig-QNM-alphah-2osqrt3} QNMs (blue spots) of the transverse gauge mode for $|\gamma|=1/12$ (the panels above are for $\gamma=1/12$ and the ones below for $\gamma=-1/12$)
and $\hat{\alpha}=2/\sqrt{3}$ in the complex frequency plane.
The left panels are the QNMs of the gauge mode, and the right ones are that of the dual gauge mode.}}
\end{figure}
\begin{figure}
\center{
\includegraphics[scale=0.28]{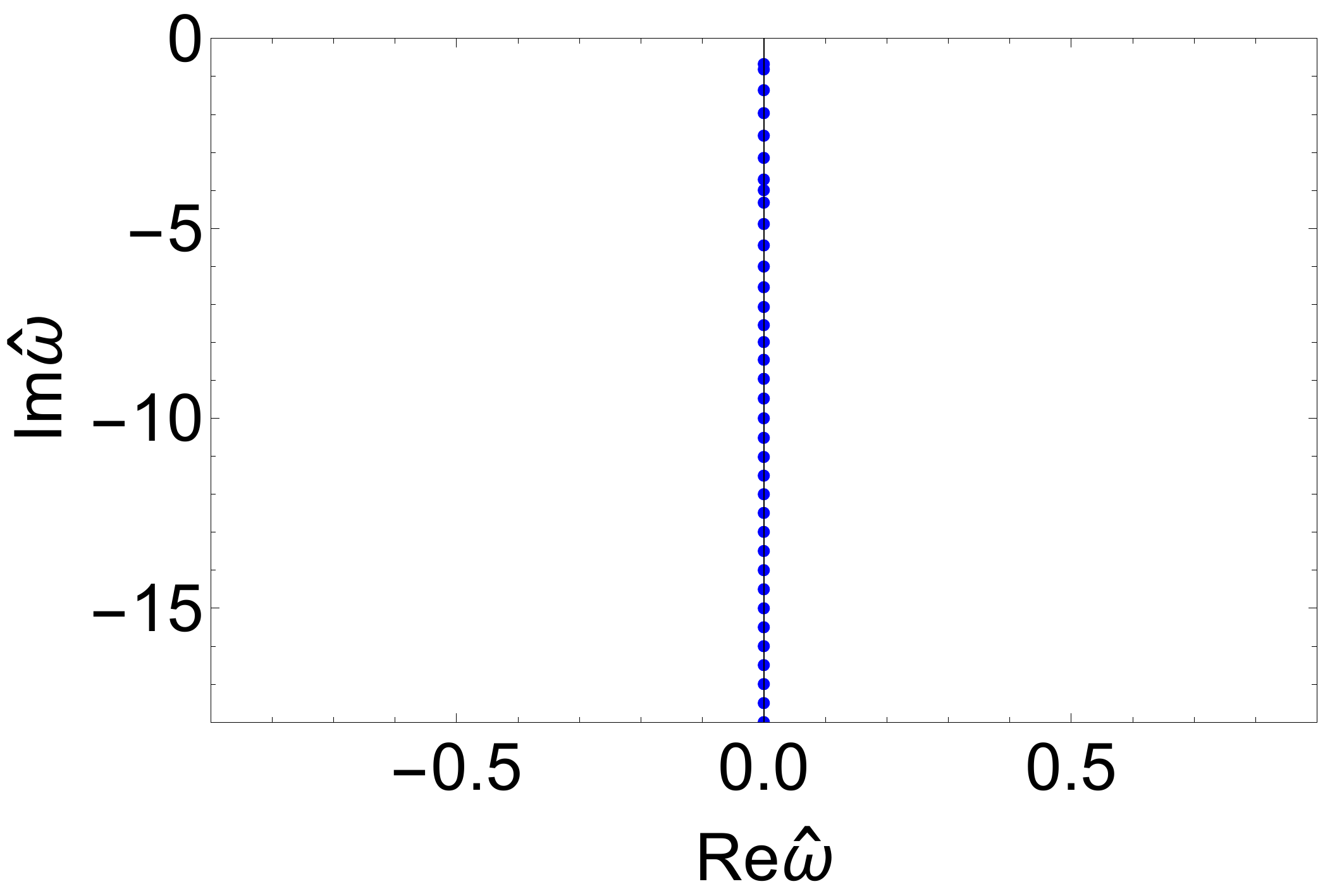}\ \hspace{0.5cm}
\includegraphics[scale=0.28]{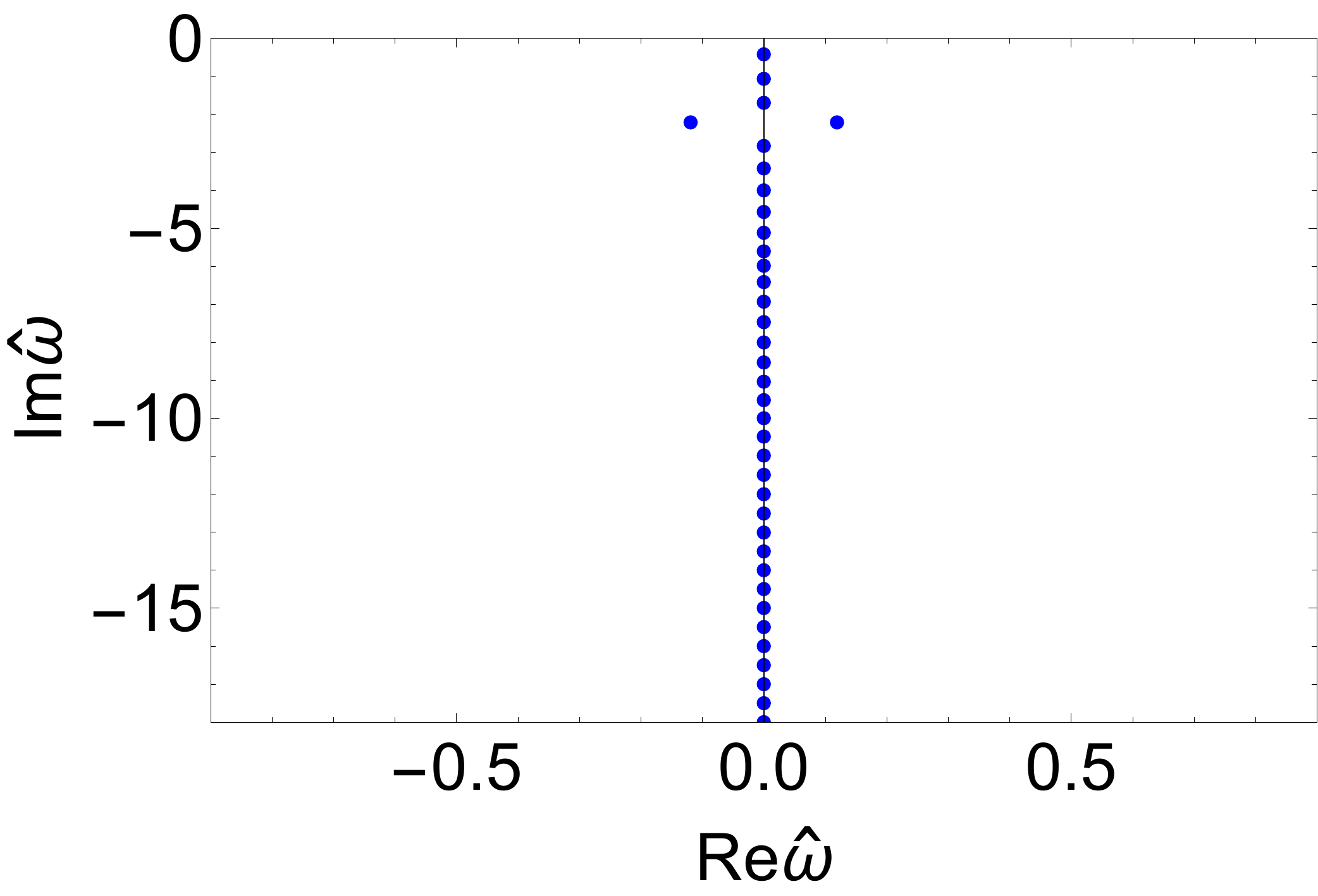}\ \\
\includegraphics[scale=0.28]{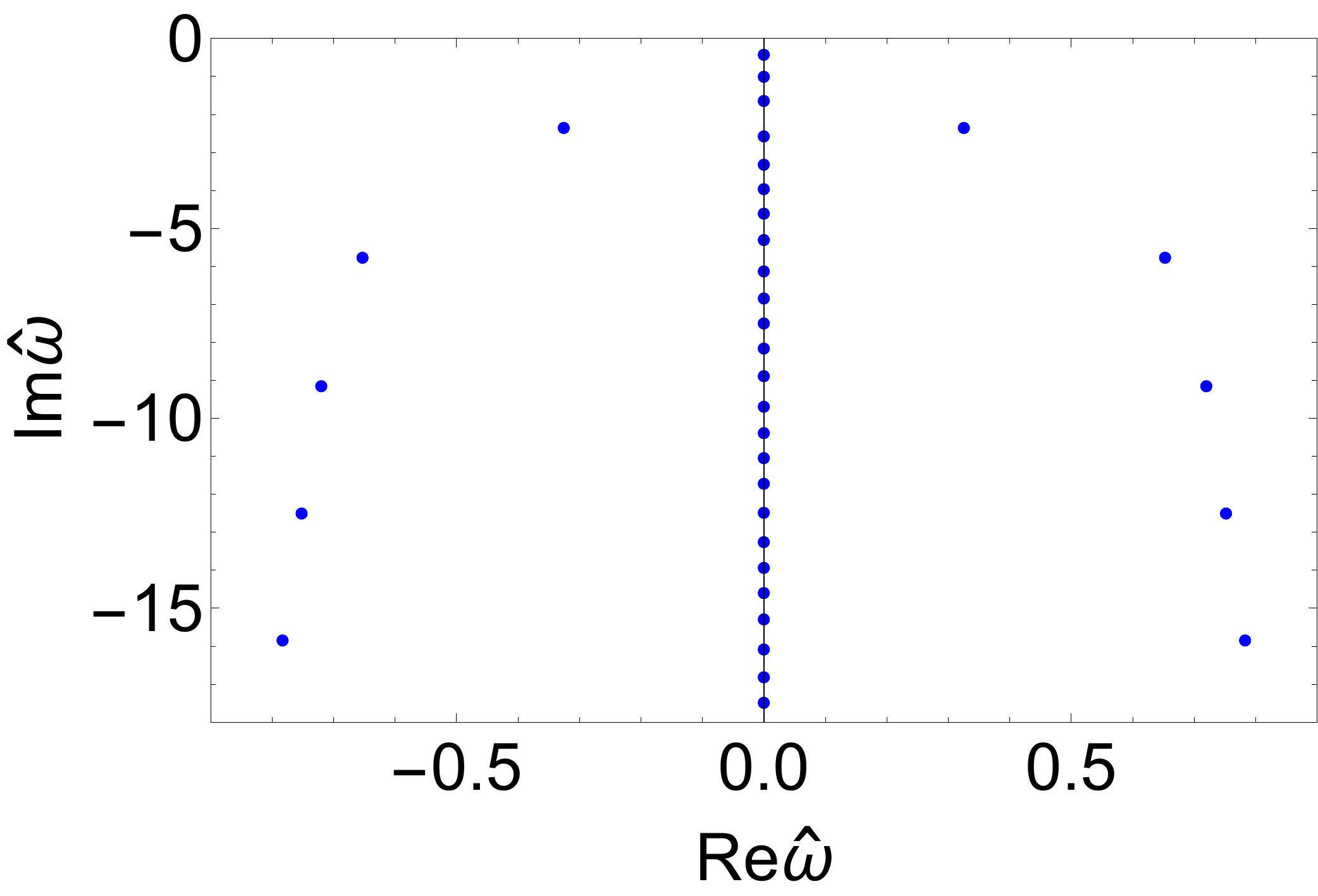}\ \hspace{0.5cm}
\includegraphics[scale=0.28]{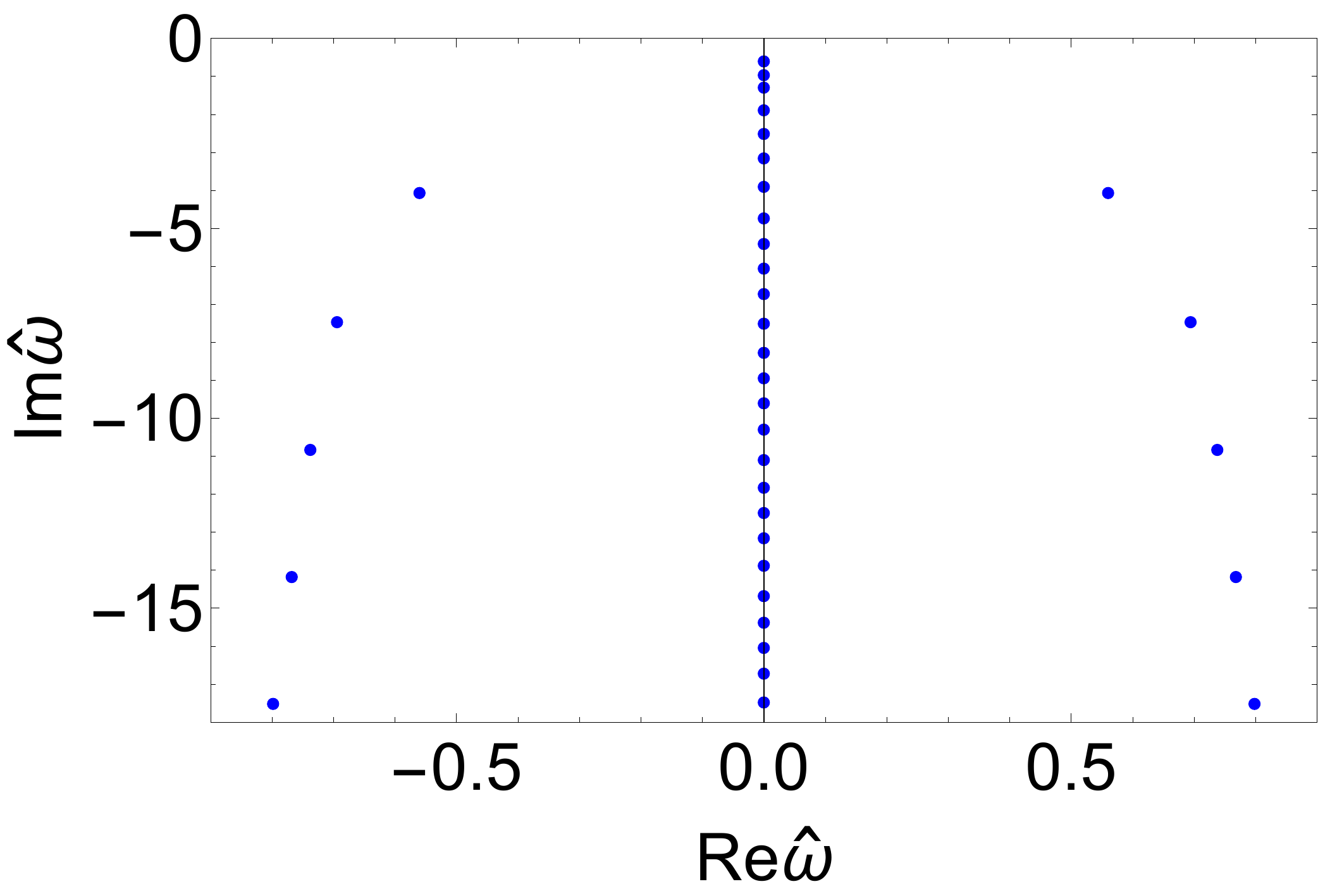}\ \\
\caption{\label{fig-QNM-alphah-2} QNMs (blue spots) of the transverse gauge mode for $|\gamma|=1/12$ (the panels above are for $\gamma=1/12$ and the ones below for $\gamma=-1/12$)
and $\hat{\alpha}=2$ in the complex frequency plane.
The left panels are the QNMs of the gauge mode, and the right ones are that of the dual gauge mode.}}
\end{figure}

FIG.\ref{fig-QNM-alphah-0}, \ref{fig-QNM-alphah-1o2}, \ref{fig-QNM-alphah-2osqrt3} and \ref{fig-QNM-alphah-2}
exhibit QNMs (blue spots) of the transverse gauge mode with representative $\hat{\alpha}$ for
$|\gamma|=1/12$ (the panels above are for $\gamma=1/12$ and the ones below for $\gamma=-1/12$)
in the complex frequency plane.
The left panels are the QNMs of the gauge mode, and the right ones are that of the dual gauge mode.
We can see that the QNMs are indeed in agreement with
the poles of the conductivity exhibited in previous section.
Since the high efficiency of numerics in searching QNMs, we exhibit more high frequency modes in FIG.\ref{fig-QNM-alphah-0}, \ref{fig-QNM-alphah-1o2}, \ref{fig-QNM-alphah-2osqrt3} and \ref{fig-QNM-alphah-2}.
They present richer information and insights on the pole structures.
We briefly present the novel observations as follows.
\begin{itemize}
  \item \textbf{Small $\hat{\alpha}$ (including $\hat{\alpha}=0$)}
  \begin{enumerate}
    \item The qualitative correspondence between the poles of $\texttt{Re}\sigma(\hat{\omega};\gamma)$
and the ones of $\texttt{Re}\sigma_{*}(\hat{\omega};-\gamma)$ approximately holds.
But we also note that with the increase of the frequency, the pole moves more quickly away from imaginary frequency axis for $\gamma<0$.
So the particle-vortex duality holds better for small frequency than for high frequency.
\item There are some poles quite away from the generic structure for $\gamma>0$. In particular, these special poles seem to emerge periodically (see the plots above in FIG.\ref{fig-QNM-alphah-0}, \ref{fig-QNM-alphah-1o2}).
  \end{enumerate}
\item \textbf{Large $\hat{\alpha}$}
\begin{enumerate}
  \item For $\gamma>0$, all of the poles of $\texttt{Re}\sigma(\hat{\omega})$
  locate on
  the imaginary frequency axis.
  But there are a pair of poles of $\texttt{Re}\sigma_{*}(\hat{\omega})$, which are off-axis.
  \item For $\gamma<0$, there are some off-axis modes at high frequency region,
  which emerges periodically. With the increase of $\hat{\alpha}$,
  the period of off-axis modes becomes large.
  \item Therefore, the correspondence between the poles of $\texttt{Re}\sigma(\hat{\omega};\gamma)$
and the ones of $\texttt{Re}\sigma_{*}(\hat{\omega};-\gamma)$ is violated.
\end{enumerate}
\end{itemize}
The above pole structures are
intriguing and are worthy of further exploration
to reveal the physics of these pole structure.

\begin{figure}
\center{
\includegraphics[scale=0.28]{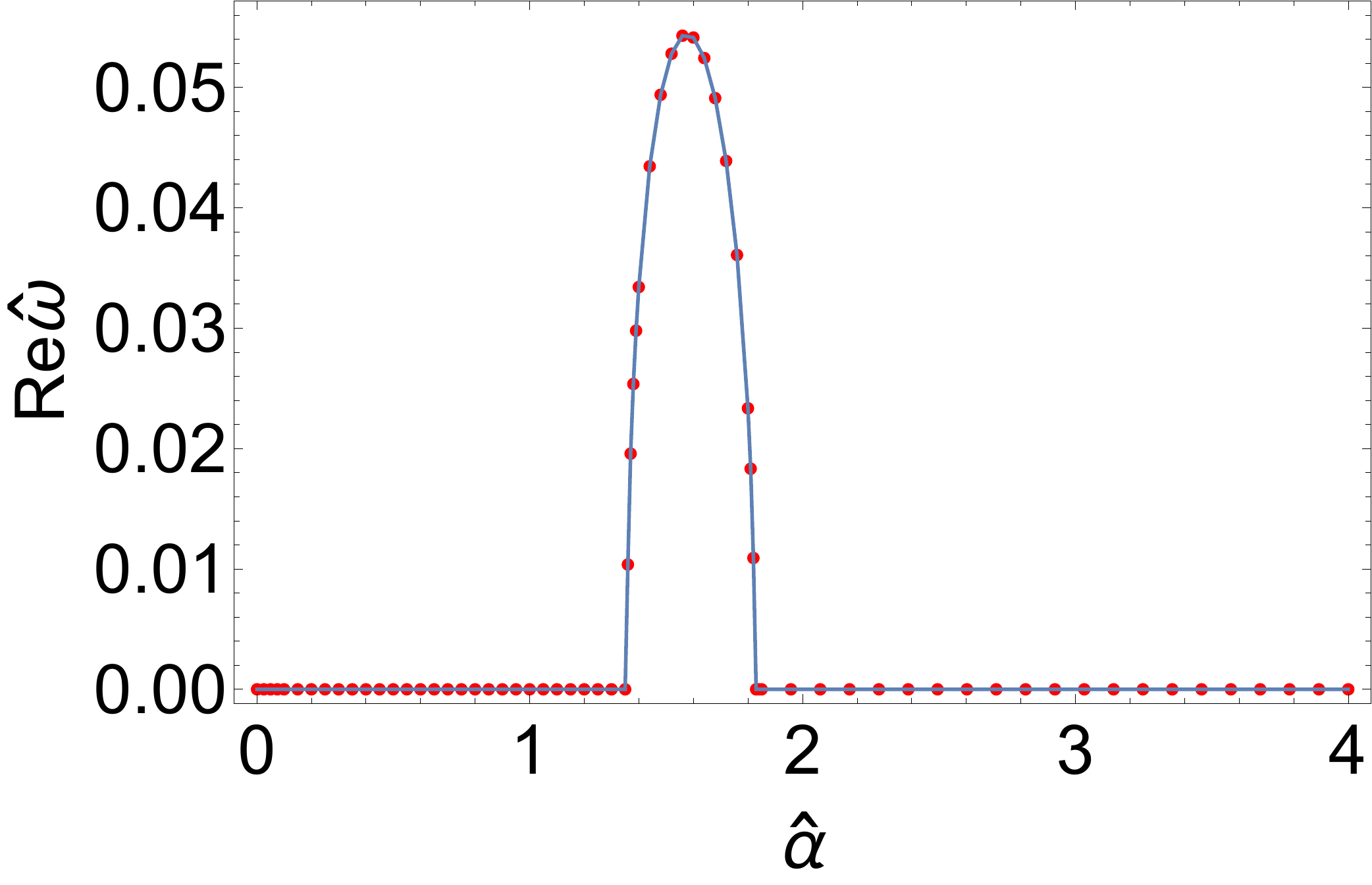}\ \hspace{0.5cm}
\includegraphics[scale=0.28]{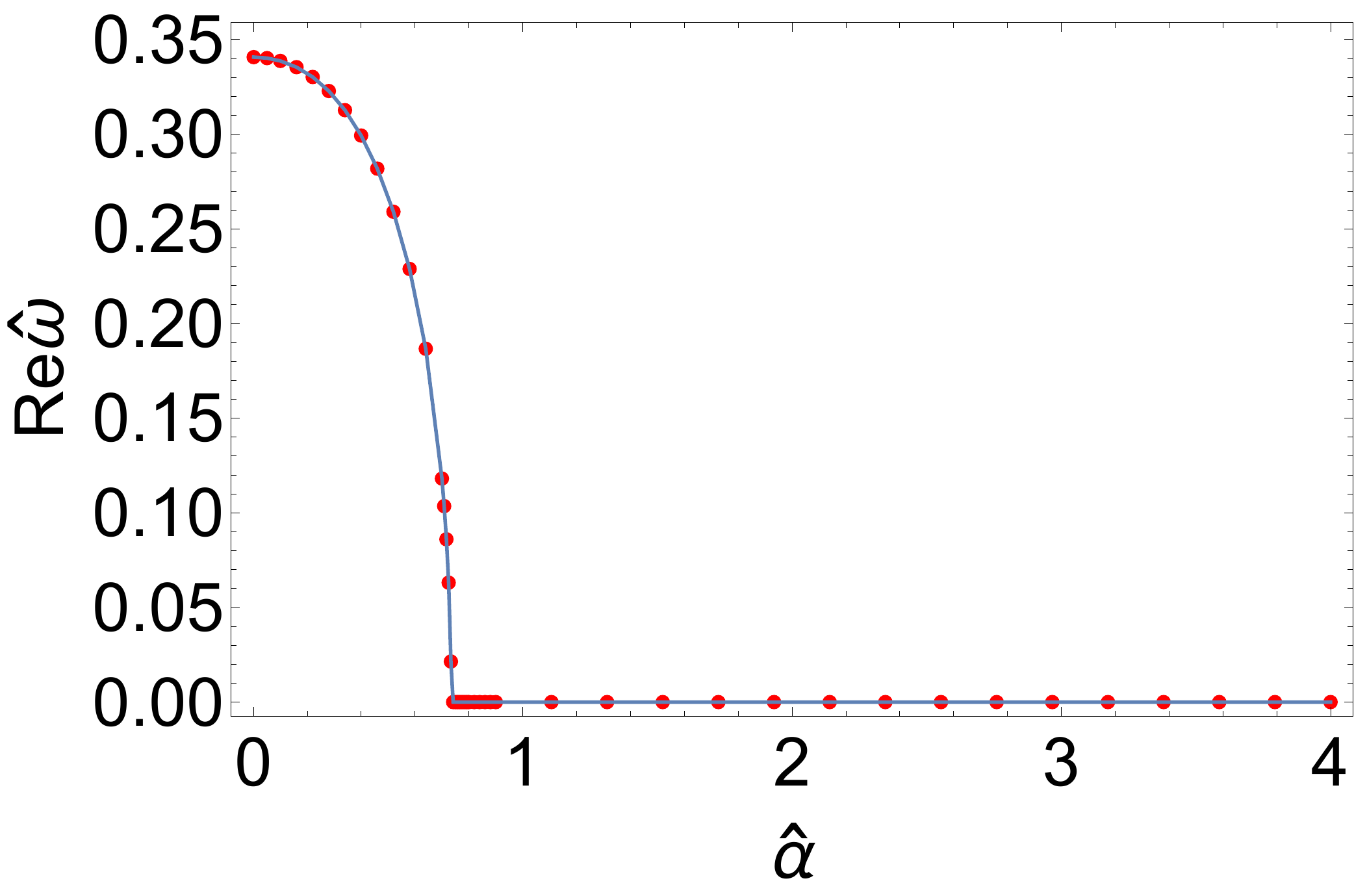}\ \\
\includegraphics[scale=0.28]{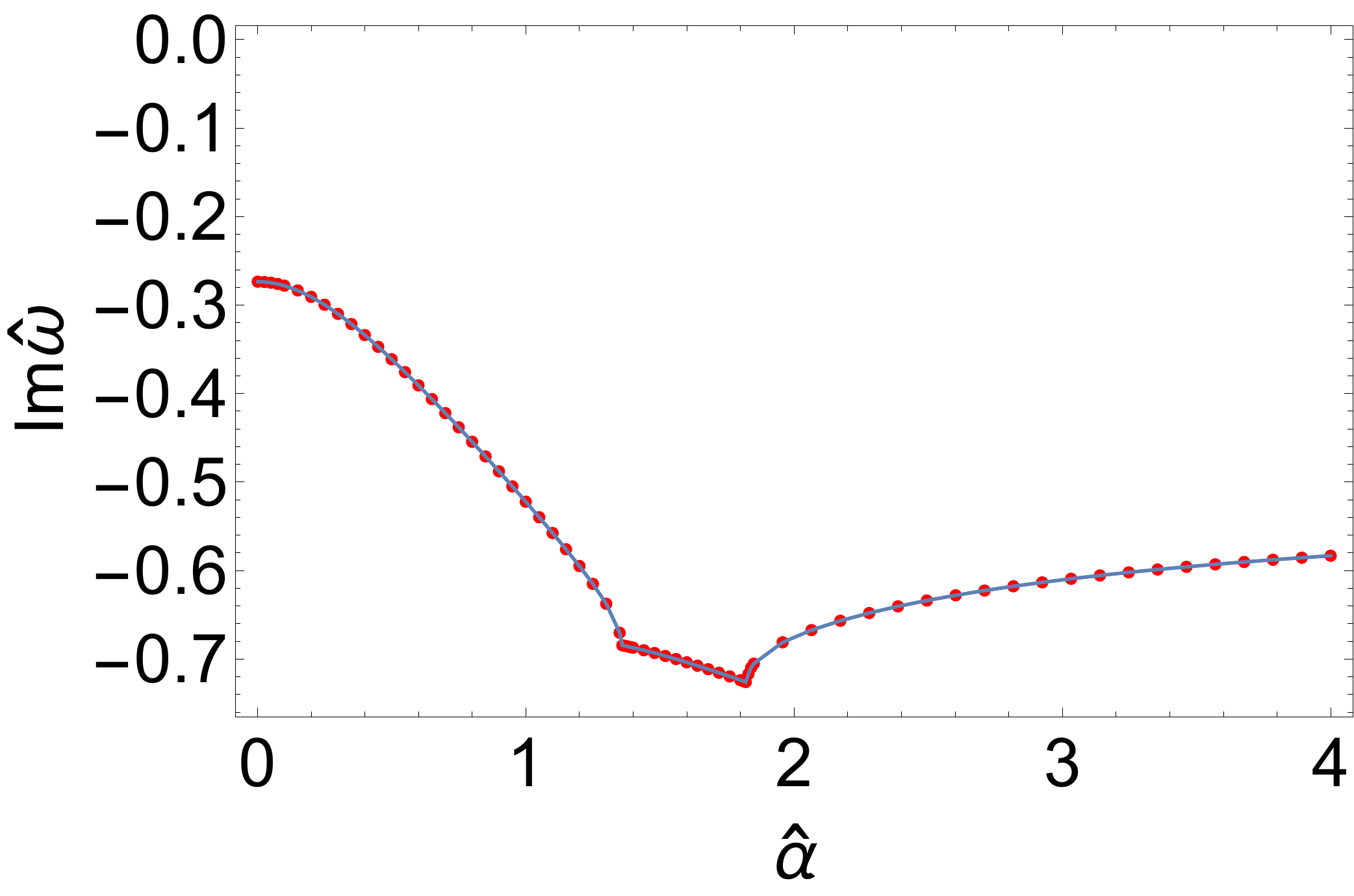}\ \hspace{0.5cm}
\includegraphics[scale=0.28]{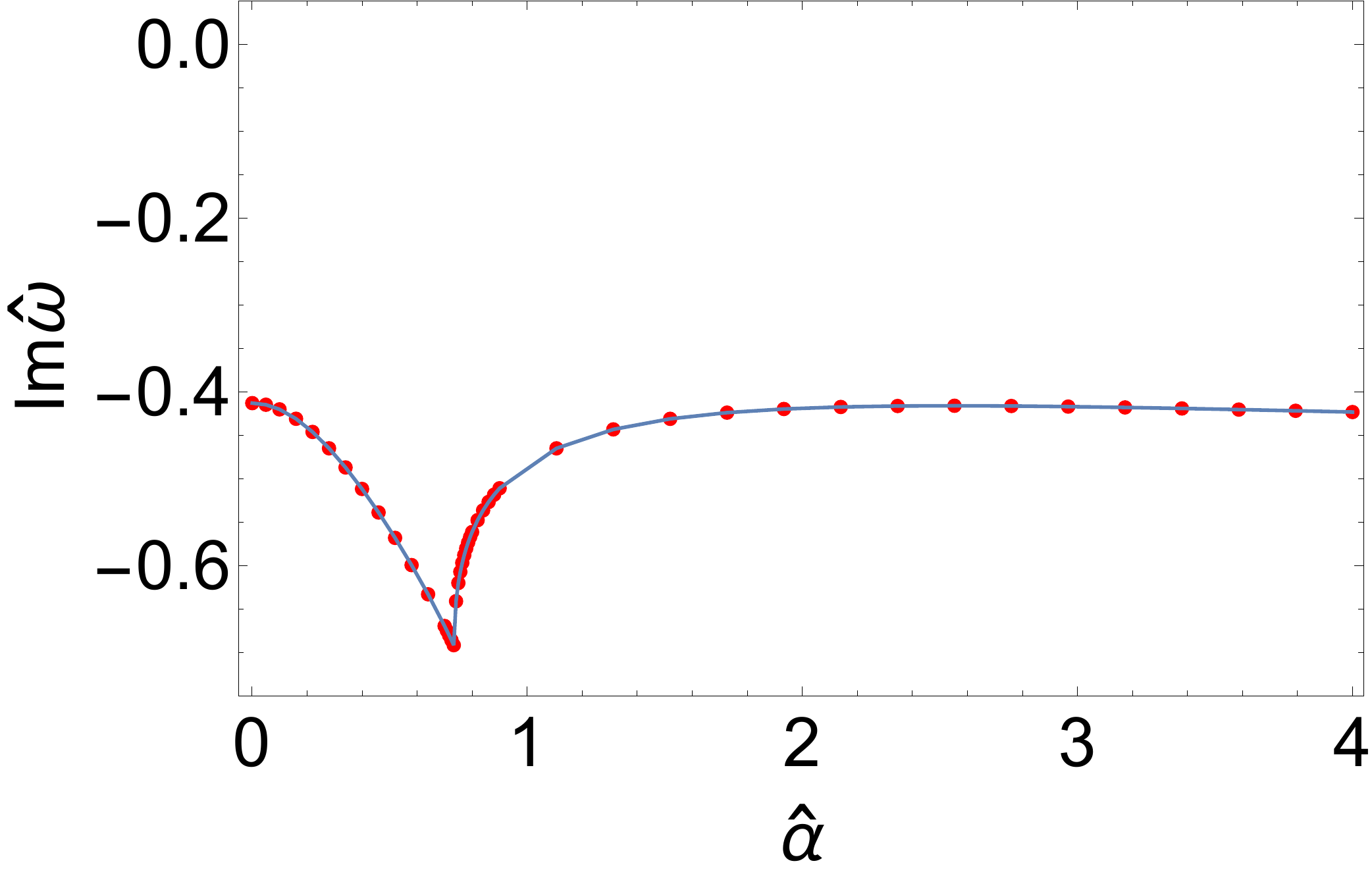}\ \\
\includegraphics[scale=0.28]{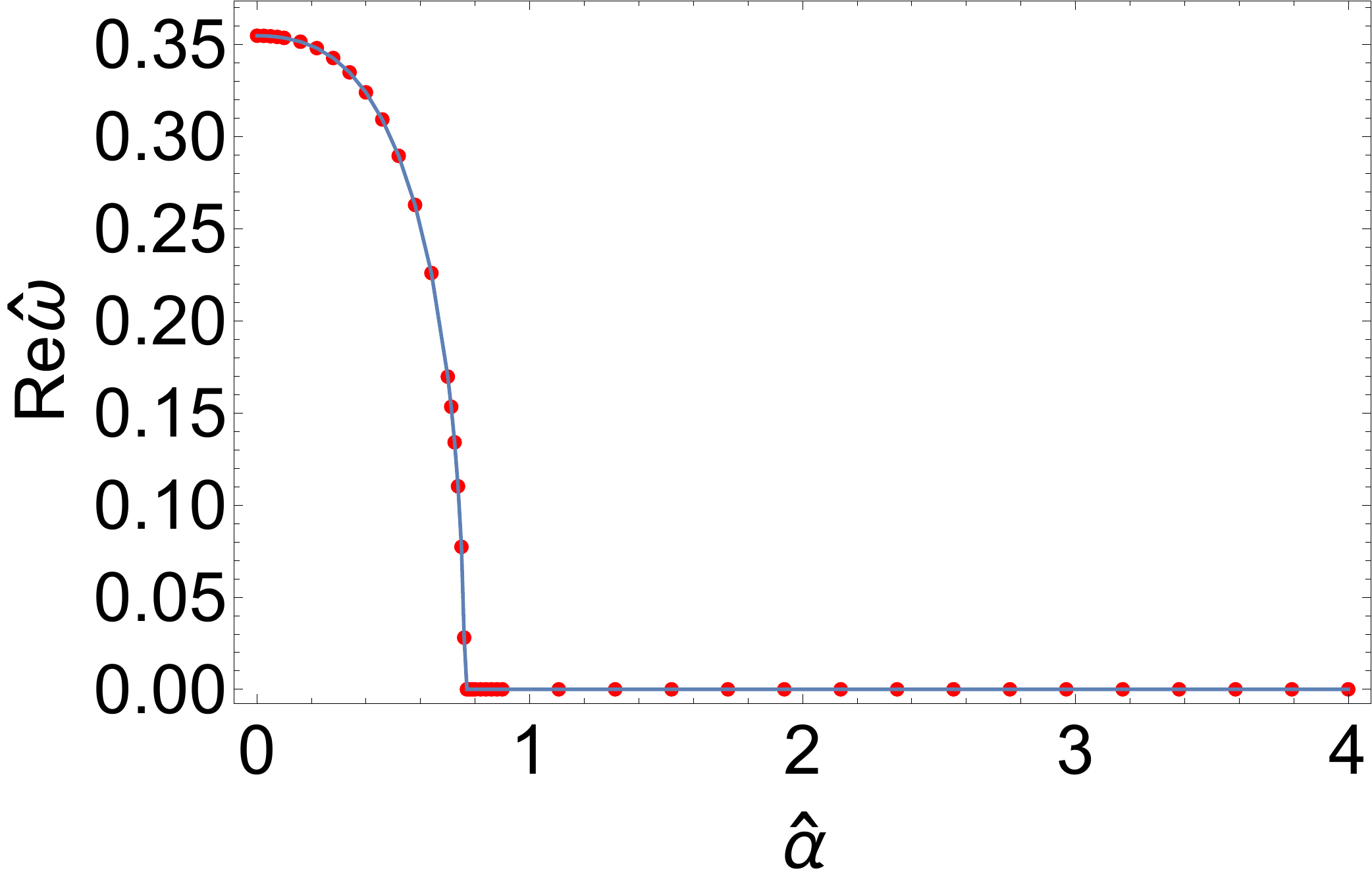}\ \hspace{0.5cm}
\includegraphics[scale=0.28]{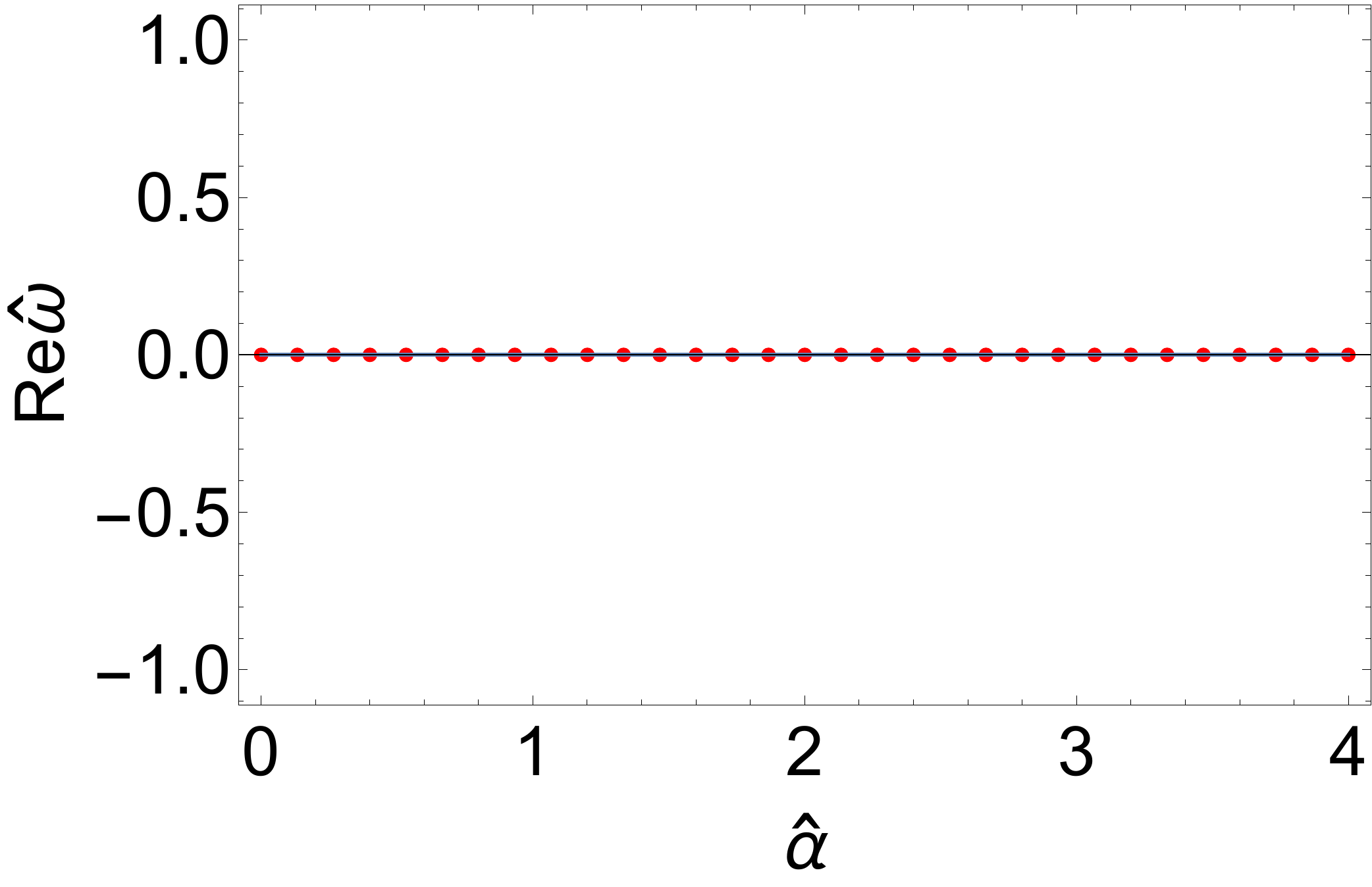}\ \\
\includegraphics[scale=0.28]{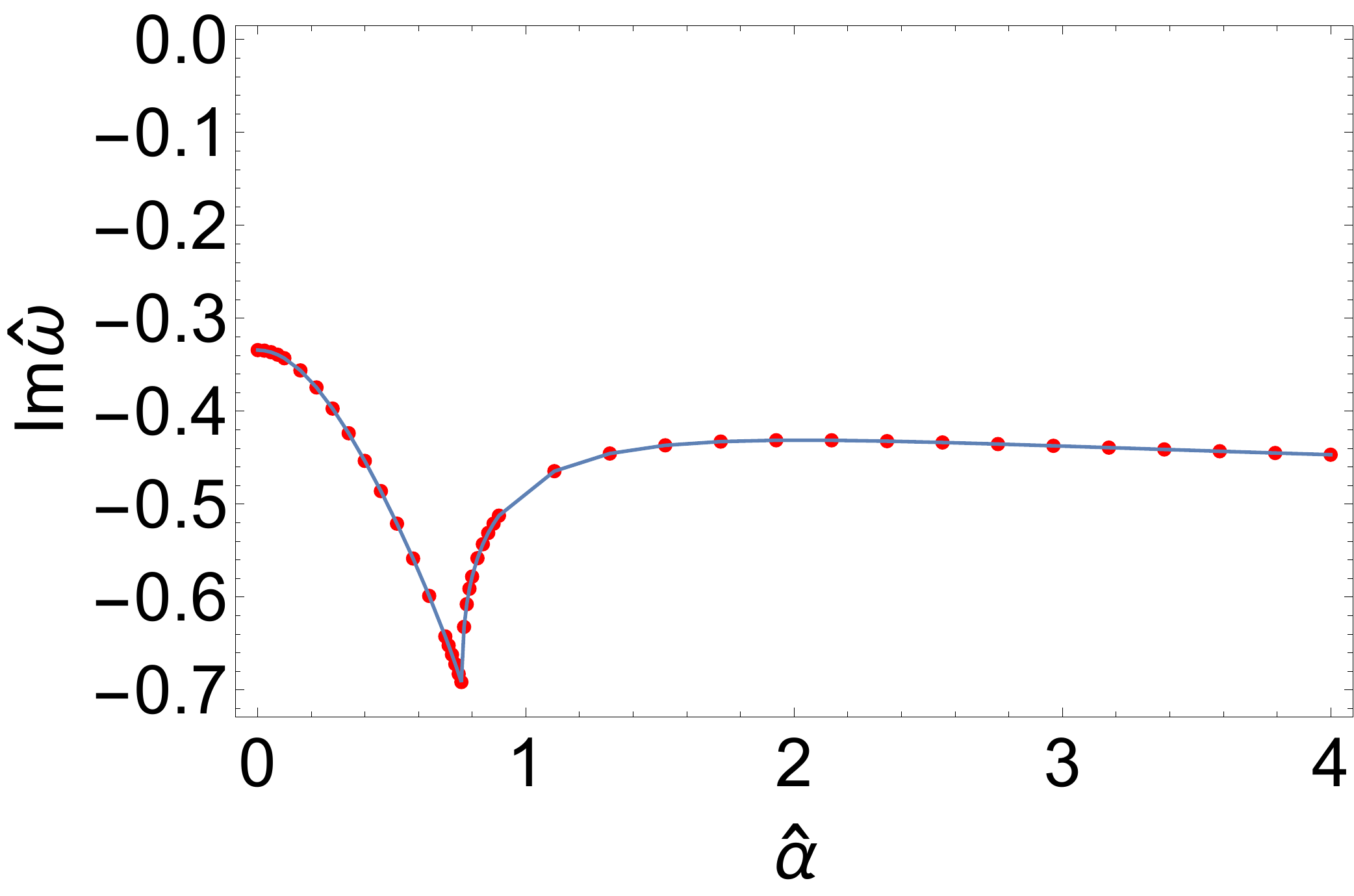}\ \hspace{0.5cm}
\includegraphics[scale=0.28]{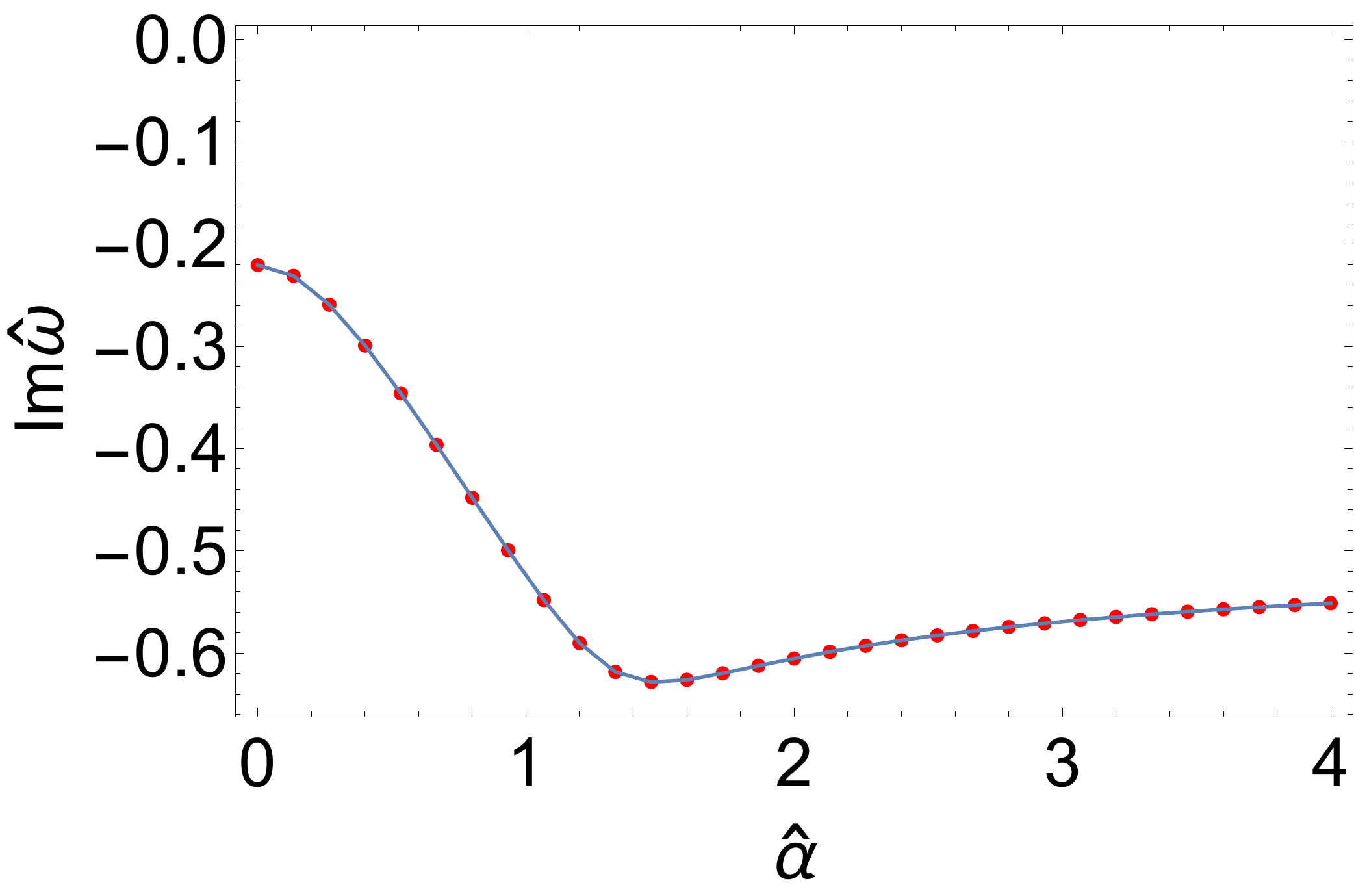}\ \\
\caption{\label{fig-QNM-vs-a} Evolution of the dominant QNM frequencies with $\hat{\alpha}$ for $|\gamma|=1/12$ (the first two rows are for $\gamma=1/12$
and the last first two ones for $\gamma=-1/12$).
The left columns are the QNM frequencies for original theory, and the right ones are that for its dual theory.}}
\end{figure}

The dominant pole describes the late time dynamical behavior of the dual system.
So, we shall deeply explore the dominant pole evolution with $\hat{\alpha}$ for $|\gamma|=1/12$ to understand the effect of the momentum dissipation.
We find that the motion of the dominant pole along the imaginary axis has the common tendency that with the increase of $\hat{\alpha}$ the pole firstly migrates downwards,
and then migrates upwards, finally approaches to a certain value.
FIG.\ref{fig-QNM-vs-a} shows this evolution.
The more detailed properties are summarized as what follows.
\begin{itemize}
  \item For $\gamma=1/12$, when the momentum dissipation is introduced,
  out of the range of $\hat{\alpha}\in(1.35,1.83)$, all the dominate poles in the original theory locate at the purely imaginary axis.
  But as revealed in \cite{Wu:2016jjd}, the low frequency conductivity in real frequency axis is a dip for large $\hat{\alpha}$.
  This gap between them calls for deeper understanding.
  We also need to understand the behavior of the QNMs belong to the region of $\hat{\alpha}\in(1.35,1.83)$, in which the pole is off-axis although the distance away from real axis is small.
  The correspondence between the poles of $\texttt{Re}\sigma(\hat{\omega};\gamma)$
and the ones of $\texttt{Re}\sigma_{*}(\hat{\omega};-\gamma)$ holds except for the range of $\hat{\alpha}\in(1.35,1.83)$.
  \item For $\gamma=-1/2$, the dominant pole in the original theory is off-axis for small $\hat{\alpha}$.
  With the increase of $\hat{\alpha}$, the poles migrate downwards and is close to the real axis.
  When $\hat{\alpha}$ reaches the value of $\hat{\alpha}\approx 0.77$,
  the poles merge into one pole locating at imaginary axis.
  After that, the pole migrates upwards along the imaginary axis as $\hat{\alpha}$ increases and approaches a fixed value for large $\hat{\alpha}$.
  This picture is consistent with the observation in \cite{Wu:2016jjd} that
  as $\hat{\alpha}$ increases the dip develops into a peak in the conductivity in real frequency axis.
  The correspondence between the poles of $\texttt{Re}\sigma(\hat{\omega};\gamma)$
and the ones of $\texttt{Re}\sigma_{*}(\hat{\omega};-\gamma)$ holds very well for almost all $\hat{\alpha}$.
\end{itemize}

\section{Conclusion and discussion}\label{sec-conclusion}

In this letter, we study the charge response in complex frequency plane and the QNMs of the boundary quantum field theory with momentum dissipation
dual to a probe generalized Maxwell system with Weyl correction.
When the strength of the momentum dissipation $\hat{\alpha}$ is small,
the pole structure of the conductivity is similar to the case without the momentum dissipation.
The qualitative correspondence between the poles of $\texttt{Re}\sigma(\hat{\omega};\gamma)$
and the ones of $\texttt{Re}\sigma_{*}(\hat{\omega};-\gamma)$ approximately holds,
which indicates that the particle-vortex duality in the dual boundary field theory.
But we also note that there are some poles quite away from the generic structure for $\gamma>0$.
In particular, these special poles seem to emerge periodically.

While the strong momentum dissipation alters the pole structure such that
most of the poles locate at the purely imaginary axis.
But there are a pair of poles of $\texttt{Re}\sigma_{*}(\hat{\omega})$ are off-axis for $\gamma>0$.
With the increase of $\hat{\alpha}$, this pair of poles migrate downwards along but close the imaginary axis.
For $\gamma<0$, there are some off-axis modes at high frequency region,
which emerges periodically. With the increase of $\hat{\alpha}$, the period of off-axis modes becomes large.
Therefore, the qualitative correspondence between the poles of $\texttt{Re}\sigma(\hat{\omega};\gamma)$
and the ones of $\texttt{Re}\sigma_{*}(\hat{\omega};-\gamma)$ is violated when the momentum dissipation is large.

We are also interested in the dominant pole.
With the increase of $\hat{\alpha}$, the dominant pole firstly migrates downwards along the imaginary axis and then migrates upwards, finally approaches to a certain value.
For $\gamma=1/12$, the correspondence between the poles of $\texttt{Re}\sigma(\hat{\omega};\gamma)$
and the ones of $\texttt{Re}\sigma_{*}(\hat{\omega};-\gamma)$ holds except for a small region of $\hat{\alpha}$.
But for $\gamma=-1/2$, the correspondence between the poles of $\texttt{Re}\sigma(\hat{\omega};\gamma)$
and the ones of $\texttt{Re}\sigma_{*}(\hat{\omega};-\gamma)$ holds very well for almost all $\hat{\alpha}$.

There are lots of questions worthy of further study.
\begin{itemize}
  \item An urgent task is to carry out an analytical study on the complex frequency conductivity on top of our present model
by matching method developed in \cite{Faulkner:2009wj},
which can provide more physical insight and understanding, in particular, for the gap between the conductivity in complex frequency and the one in real frequency observed for $\gamma=1/12$.
\item In this letter, we only study the QNMs at zero density. It would be worth to consider the model at finite charge density and compute the QNMs beyond the probe limit.
But when we take into account the full backreaction of our present model, the equations of motion become a set of third order differential equations
with high nonlinearity, which is very hard to solve analytically so far.
The nonlinearity problem could be attacked in the small limit of $\gamma$ by expanding the equations of motions up to the first order of $\gamma$. However, even if
we have obtained the perturbative black brane solution to the first order of $\gamma$ in \cite{Ling:2016dck},
we still need to solve the linear perturbative differential equations beyond the second order to obtain the QNMs.
It is a hard task and so we shall deeply explore it in future.
\item We can also study the QNMs of our present model with finite momentum, which reveals richer physics of the system.
\item The superconducting phase with Weyl correction has been widely explored in \cite{Wu:2010vr,Ma:2011zze,Wu:2017xki,Momeni:2011ca,Momeni:2012ab,Zhao:2012kp,Momeni:2013fma,Momeni:2014efa,Zhang:2015eea,Mansoori:2016zbp}
and references therein. It is also interesting to explore the QNMs in the superconducting phase of these models.
\end{itemize}

\begin{acknowledgments}

This work is supported by the Natural Science Foundation of China under
Grant Nos. 11775036 and 11305018, and by Natural Science Foundation of Liaoning Province under
Grant No.201602013.

\end{acknowledgments}

\begin{appendix}

\end{appendix}


\begin{thebibliography}{99}


\bibitem{Maldacena:1997re}
  J.~M.~Maldacena,
  Int.\ J.\ Theor.\ Phys.\  {\bf 38}, 1113 (1999)
  [Adv.\ Theor.\ Math.\ Phys.\  {\bf 2}, 231 (1998)]
  [hep-th/9711200].

\bibitem{Gubser:1998bc}
  S.~S.~Gubser, I.~R.~Klebanov and A.~M.~Polyakov,
  Phys.\ Lett.\ B {\bf 428}, 105 (1998)
  [hep-th/9802109].

\bibitem{Witten:1998qj}
  E.~Witten,
  Adv.\ Theor.\ Math.\ Phys.\  {\bf 2}, 253 (1998)
  [hep-th/9802150].

\bibitem{Aharony:1999ti}
  O.~Aharony, S.~S.~Gubser, J.~M.~Maldacena, H.~Ooguri and Y.~Oz,
  Phys.\ Rept.\  {\bf 323}, 183 (2000)
  [hep-th/9905111].

\bibitem{Myers:2010pk}
  R.~C.~Myers, S.~Sachdev and A.~Singh,
  Phys.\ Rev.\ D {\bf 83}, 066017 (2011)
  [arXiv:1010.0443 [hep-th]].

\bibitem{Sachdev:2011wg}
  S.~Sachdev,
  Ann.\ Rev.\ Condensed Matter Phys.\  {\bf 3}, 9 (2012)
  [arXiv:1108.1197 [cond-mat.str-el]].

\bibitem{Hartnoll:2016apf}
  S.~A.~Hartnoll, A.~Lucas and S.~Sachdev,
  arXiv:1612.07324 [hep-th].


\bibitem{Ritz:2008kh}
  A.~Ritz and J.~Ward,
  Phys.\ Rev.\ D {\bf 79}, 066003 (2009)
  [arXiv:0811.4195 [hep-th]].

\bibitem{WitczakKrempa:2012gn}
  W.~Witczak-Krempa and S.~Sachdev,
  Phys.\ Rev.\ B {\bf 86}, 235115 (2012)
  [arXiv:1210.4166 [cond-mat.str-el]].

\bibitem{WitczakKrempa:2013ht}
  W.~Witczak-Krempa and S.~Sachdev,
  Phys.\ Rev.\ B {\bf 87}, 155149 (2013)
  [arXiv:1302.0847 [cond-mat.str-el]].

\bibitem{Witczak-Krempa:2013nua}
  W.~Witczak-Krempa, E.~S.~S{\o}rensen and S.~Sachdev,
  Nature Phys.\  {\bf 10}, 361 (2014)
  [arXiv:1309.2941 [cond-mat.str-el]].

\bibitem{Katz:2014rla}
  E.~Katz, S.~Sachdev, E.~S.~S{\o}rensen and W.~Witczak-Krempa,
  Phys.\ Rev.\ B {\bf 90}, no. 24, 245109 (2014)
  [arXiv:1409.3841 [cond-mat.str-el]].


\bibitem{Damle:1997rxu}
  K.~Damle and S.~Sachdev,
  Phys.\ Rev.\ B {\bf 56}, no. 14, 8714 (1997)
  [cond-mat/9705206 [cond-mat.str-el]].

\bibitem{Chen:2017dsy}
  C.~F.~Chen and A.~Lucas,
  Phys.\ Lett.\ B {\bf 774}, 569 (2017)
  [arXiv:1709.01520 [hep-th]].

\bibitem{Baggioli:2016oju}
  M.~Baggioli and O.~Pujolas,
  JHEP {\bf 1612}, 107 (2016)
  [arXiv:1604.08915 [hep-th]].


\bibitem{Witczak-Krempa:2013aea}
  W.~Witczak-Krempa,
  Phys.\ Rev.\ B {\bf 89}, no. 16, 161114 (2014)
  [arXiv:1312.3334 [cond-mat.str-el]].

\bibitem{Myers:2016wsu}
  R.~C.~Myers, T.~Sierens and W.~Witczak-Krempa,
  JHEP {\bf 1605}, 073 (2016)
  Addendum: [JHEP {\bf 1609}, 066 (2016)]
  [arXiv:1602.05599 [hep-th]].

\bibitem{Lucas:2017dqa}
  A.~Lucas, T.~Sierens and W.~Witczak-Krempa,
  JHEP {\bf 1707}, 149 (2017)
  [arXiv:1704.05461 [hep-th]].

\bibitem{Wu:2016jjd}
  J.~P.~Wu,
  arXiv:1609.04729 [hep-th].

\bibitem{Fu:2017oqa}
  G.~Fu, J.~P.~Wu, B.~Xu and J.~Liu,
  Phys.\ Lett.\ B {\bf 769}, 569 (2017)
  [arXiv:1705.06672 [hep-th]].


\bibitem{Kovtun:2005ev}
  P.~K.~Kovtun and A.~O.~Starinets,
  Phys.\ Rev.\ D {\bf 72}, 086009 (2005)
  [hep-th/0506184].

\bibitem{Berti:2009kk}
  E.~Berti, V.~Cardoso and A.~O.~Starinets,
  Class.\ Quant.\ Grav.\  {\bf 26}, 163001 (2009)
  [arXiv:0905.2975 [gr-qc]].

\bibitem{Konoplya:2011qq}
  R.~A.~Konoplya and A.~Zhidenko,
  Rev.\ Mod.\ Phys.\  {\bf 83}, 793 (2011)
  [arXiv:1102.4014 [gr-qc]].

\bibitem{Horowitz:1999jd}
  G.~T.~Horowitz and V.~E.~Hubeny,
  Phys.\ Rev.\ D {\bf 62}, 024027 (2000)
  [hep-th/9909056].

\bibitem{Wang:2004bv}
  B.~Wang, C.~Y.~Lin and C.~Molina,
  Phys.\ Rev.\ D {\bf 70}, 064025 (2004)
  [hep-th/0407024].

\bibitem{Wang:2000gsa}
  B.~Wang, C.~Y.~Lin and E.~Abdalla,
  Phys.\ Lett.\ B {\bf 481}, 79 (2000)
  [hep-th/0003295].

\bibitem{Giammatteo:2004wp}
  M.~Giammatteo and J.~l.~Jing,
  Phys.\ Rev.\ D {\bf 71}, 024007 (2005)
  [gr-qc/0403030].

\bibitem{Yao:2011kf}
  W.~Yao, S.~Chen and J.~Jing,
  Phys.\ Rev.\ D {\bf 83}, 124018 (2011)
  [arXiv:1101.0042 [gr-qc]].

\bibitem{Lin:2016sch}
  K.~Lin and W.~L.~Qian,
  Class.\ Quant.\ Grav.\  {\bf 34}, no. 9, 095004 (2017)
  [arXiv:1610.08135 [gr-qc]].

\bibitem{Gursoy:2016ggq}
  U.~G\"ursoy, A.~Jansen and W.~van der Schee,
  Phys.\ Rev.\ D {\bf 94}, no. 6, 061901 (2016)
  [arXiv:1603.07724 [hep-th]].

\bibitem{Gursoy:2016tgf}
  U.~G\"ursoy, A.~Jansen, W.~Sybesma and S.~Vandoren,
  Phys.\ Rev.\ Lett.\  {\bf 117}, no. 5, 051601 (2016)
  [arXiv:1602.01375 [hep-th]].

\bibitem{Jansen:2017oag}
  A.~Jansen,
  arXiv:1709.09178 [gr-qc].

\bibitem{Cardoso:2017soq}
  V.~Cardoso, J.~L.~Costa, K.~Destounis, P.~Hintz and A.~Jansen,
  arXiv:1711.10502 [gr-qc]

\bibitem{Andrade:2013gsa}
  T.~Andrade and B.~Withers,
  JHEP {\bf 1405}, 101 (2014)
  [arXiv:1311.5157 [hep-th]].

\bibitem{Davison:2014lua}
  R.~A.~Davison and B.~Gouteraux,
  JHEP {\bf 1501}, 039 (2015)
  [arXiv:1411.1062 [hep-th]].

\bibitem{Grozdanov:2015qia}
  S.~Grozdanov, A.~Lucas, S.~Sachdev and K.~Schalm,
  Phys.\ Rev.\ Lett.\  {\bf 115}, no. 22, 221601 (2015)
  [arXiv:1507.00003 [hep-th]].

\bibitem{Baggioli:2014roa}
  M.~Baggioli and O.~Pujolas,
  Phys.\ Rev.\ Lett.\  {\bf 114}, no. 25, 251602 (2015)
  [arXiv:1411.1003 [hep-th]].

\bibitem{Kuang:2017cgt}
  X.~M.~Kuang and J.~P.~Wu,
  Phys.\ Lett.\ B {\bf 770}, 117 (2017)
  [arXiv:1702.01490 [hep-th]].

\bibitem{Ge:2014aza}
  X.~H.~Ge, Y.~Ling, C.~Niu and S.~J.~Sin,
  Phys.\ Rev.\ D {\bf 92}, no. 10, 106005 (2015)
  [arXiv:1412.8346 [hep-th]].

\bibitem{Ling:2016dck}
  Y.~Ling, P.~Liu, J.~P.~Wu and Z.~Zhou,
  Phys.\ Lett.\ B {\bf 766}, 41 (2017)
  [arXiv:1606.07866 [hep-th]].

\bibitem{Kim:2014bza}
  K.~Y.~Kim, K.~K.~Kim, Y.~Seo and S.~J.~Sin,
  JHEP {\bf 1412}, 170 (2014)
  [arXiv:1409.8346 [hep-th]].

\bibitem{Nicolis:2015sra}
  A.~Nicolis, R.~Penco, F.~Piazza and R.~Rattazzi,
  JHEP {\bf 1506}, 155 (2015)
  [arXiv:1501.03845 [hep-th]].

\bibitem{Alberte:2015isw}
  L.~Alberte, M.~Baggioli, A.~Khmelnitsky and O.~Pujolas,
  JHEP {\bf 1602}, 114 (2016)
  [arXiv:1510.09089 [hep-th]].

\bibitem{Faulkner:2009wj}
  T.~Faulkner, H.~Liu, J.~McGreevy and D.~Vegh,
  Phys.\ Rev.\ D {\bf 83}, 125002 (2011)
  [arXiv:0907.2694 [hep-th]].

\bibitem{Wu:2010vr}
  J.~P.~Wu, Y.~Cao, X.~M.~Kuang and W.~J.~Li,
  Phys.\ Lett.\ B {\bf 697}, 153 (2011)
  [arXiv:1010.1929 [hep-th]].

\bibitem{Ma:2011zze}
  D.~Z.~Ma, Y.~Cao and J.~P.~Wu,
  Phys.\ Lett.\ B {\bf 704}, 604 (2011)
  [arXiv:1201.2486 [hep-th]].

\bibitem{Wu:2017xki}
  J.~P.~Wu and P.~Liu,
  Phys.\ Lett.\ B {\bf 774}, 527 (2017)
  [arXiv:1710.07971 [hep-th]].

\bibitem{Momeni:2011ca}
  D.~Momeni and M.~R.~Setare,
  Mod.\ Phys.\ Lett.\ A {\bf 26}, 2889 (2011)
  [arXiv:1106.0431 [physics.gen-ph]].
\bibitem{Momeni:2012ab}
  D.~Momeni, N.~Majd and R.~Myrzakulov,
  Europhys.\ Lett.\  {\bf 97}, 61001 (2012)
  [arXiv:1204.1246 [hep-th]].
\bibitem{Zhao:2012kp}
  Z.~Zhao, Q.~Pan and J.~Jing,
  Phys.\ Lett.\ B {\bf 719}, 440 (2013)
  [arXiv:1212.3062].
\bibitem{Momeni:2013fma}
  D.~Momeni, R.~Myrzakulov and M.~Raza,
  Int.\ J.\ Mod.\ Phys.\ A {\bf 28}, 1350096 (2013)
  [arXiv:1307.8348 [hep-th]].
\bibitem{Momeni:2014efa}
  D.~Momeni, M.~Raza and R.~Myrzakulov,
  Int.\ J.\ Geom.\ Meth.\ Mod.\ Phys.\  {\bf 13}, 1550131 (2016)
  [arXiv:1410.8379 [hep-th]].
\bibitem{Zhang:2015eea}
  L.~Zhang, Q.~Pan and J.~Jing,
  Phys.\ Lett.\ B {\bf 743}, 104 (2015)
  [arXiv:1502.05635 [hep-th]].

\bibitem{Mansoori:2016zbp}
  S.~A.~H.~Mansoori, B.~Mirza, A.~Mokhtari, F.~L.~Dezaki and Z.~Sherkatghanad,
  JHEP {\bf 1607}, 111 (2016)
  [arXiv:1602.07245 [hep-th]].




\end{thebibliography}
\end{document}